\documentclass[10pt]{article}
\usepackage{ctex} 
\usepackage{newtxtext,newtxmath}

\usepackage{graphicx}

\usepackage[letterpaper,margin=1in]{geometry}

\renewenvironment{abstract}
	{\quotation}
	{\endquotation}

\date{}

\makeatletter
\renewcommand{\fnum@figure}{\textbf{Figure \thefigure}}
\renewcommand{\fnum@table}{\textbf{Table \thetable}}
\makeatother

\usepackage{scicite}
\usepackage{multicol}
\usepackage{caption}
\usepackage{url}
\usepackage{hyperref}

\def\scititle{
Super-earths and mini-neptunes follow different orbital period--eccentricity relations
}
\title{\bfseries \boldmath \scititle}

\author{
	Ke-Ting Shin (辛科霆)$^{1,2\dagger}$,
	Dong-Sheng An (安东升)$^{1,2\dagger}$,
	Ji-Wei Xie (谢基伟)$^{1,2\ast}$,\and
        Ji-Lin Zhou (周济林)$^{1,2}$,
        Fei Dai (戴飞)$^{3}$\and
	\small$^{1}$School of Astronomy and Space Science, Nanjing University, Nanjing, China.\\
	\small$^{2}$Key Laboratory of Modern Astronomy and Astrophysics, Ministry of Education, Nanjing, China.\\
        \small$^{3}$Institute for Astronomy, University of Hawai`i, Honolulu, HI, USA.\\
    \small$^\dagger$These authors contributed equally to this work.\\
	\small$^\ast$Corresponding author. Email: jwxie@nju.edu.cn
}

\begin{document} 

\maketitle

\begin{abstract} \fontsize{13pt}{15pt} \selectfont \bfseries \boldmath
Many exoplanets have been observed with radius sizes between that of Earth and that of Neptune and are thus classified into two groups: super-earths (SEs) and mini-neptunes (MNs).
There are no SEs and MNs in the Solar System, and the mechanisms responsible for their formation and evolution are debated. 
We investigate the relationships between the orbital period and eccentricity of SEs and MNs using both ensemble analyses and individual measurements.
We found that MNs follow an anti-correlation between orbital period and eccentricity, but SEs follow a different relation, possibly in the opposite direction.
These trends imply that MNs and SEs are dynamically distinct populations.
We suggest that SEs have been more strongly influenced by violent processes such as gravitational scattering and giant impacts, whereas MNs predominantly experienced quiescent secular evolution.
\end{abstract}

\clearpage
\begin{multicols}{2} 
\noindent The eccentricity of a celestial body's orbit can quantify its dynamic state.
For a planet or exoplanet, the eccentricity affects the amount of radiation it receives from its host star and therefore its climate and habitability\cite{2002IJAsB...1...61W,2010ApJ...721.1295D,2016A&A...591A.106B}.
For a population of exoplanets, the distribution of orbital eccentricity and its dependence on other parameters are determined by the formation and evolution mechanisms\cite{2013ApJ...767L..24D,2015PNAS..112...20L,2023PNAS..12017398S}.
For giant exoplanets, the relationship between orbital period ($P$) and eccentricity ($e$) has been used to investigate their evolution\cite{2005A&A...431.1129H,2018ARA&A..56..175D}, including star-planet tidal interactions\cite{2008ApJ...678.1396J}, planet-planet interactions\cite{2008ApJ...686..621F,2008ApJ...686..580C}, and interactions between planets and the protoplanetary disks from which they formed\cite{2003ApJ...585.1024G,2013ApJ...769...26C}.

For smaller exoplanets, the relationship(s) between $P$ and $e$ is less clear.
Previous studies have reported a flat $P$--$e$ distribution\cite{2020A&A...635A..37C,2024AJ....168..115B}, unlike the positive correlation found for giant exoplanets\cite{2005A&A...431.1129H,2024AJ....168..115B}.
Those studies assumed that all exoplanets smaller than a gas giant form a single population, but observations have shown that those exoplanets have a bimodal radius distribution, interpreted as rocky super-earths (SEs) and gaseous mini-neptunes (MNs) separated by a radius valley\cite{2017AJ....154..109F}.
These are known to be physically distinct classes of exoplanet with different compositions, as indicated by their different densities\cite{2019PNAS..116.9723Z,2022Sci...377.1211L}.
We therefore investigated the $P$--$e$ relations of MNs and SEs as separate populations.

\subsubsection*{Statistical analysis of transiting exoplanets}
\begin{figure}[!ht]
\centering
\includegraphics[width=1\textwidth]{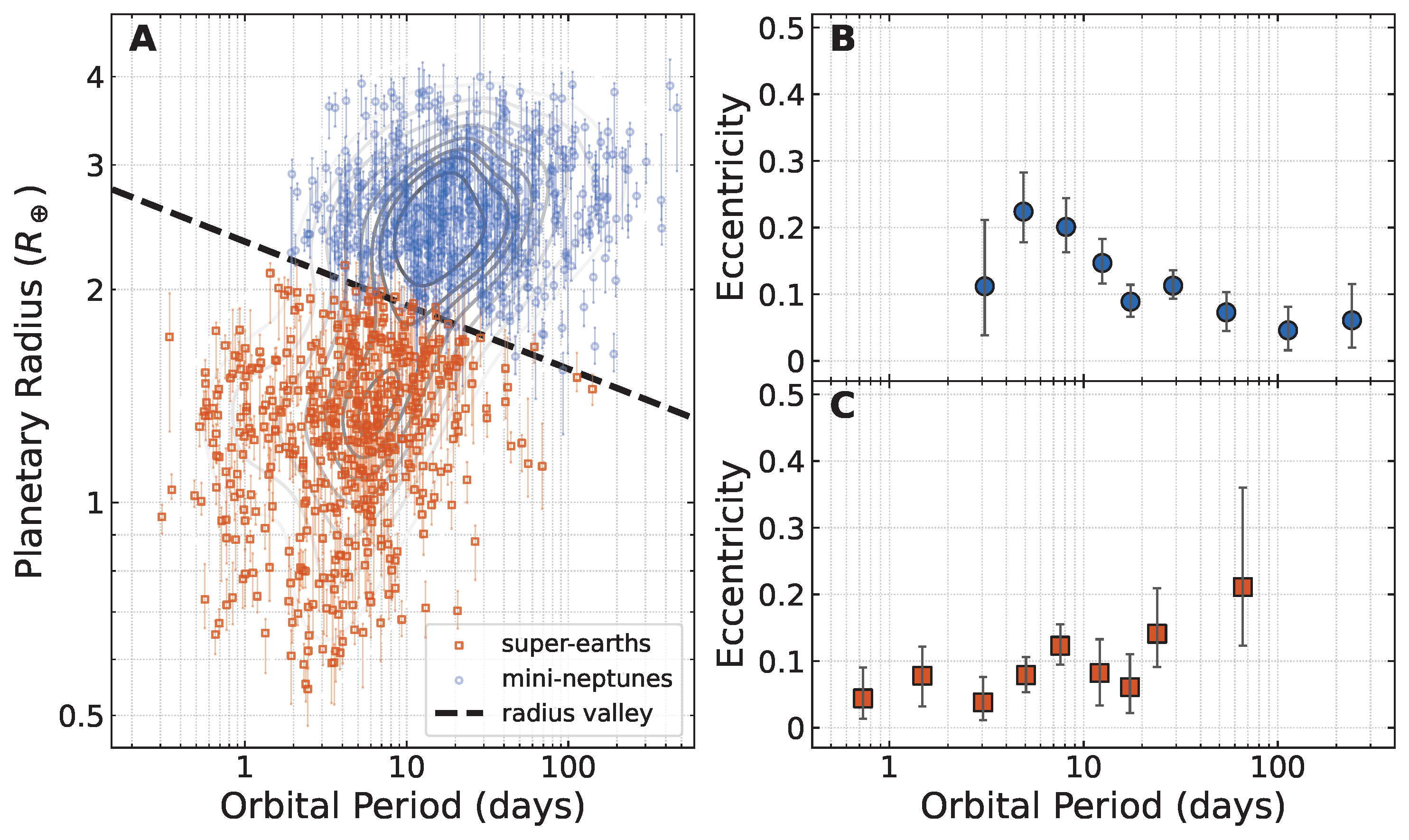}
\caption{
\textbf{Radii and eccentricities of the fiducial Kepler sample.}
{\bf (A)} Planet radius (in earth radii, $R_\oplus$) as a function of orbital period for SEs (orange squares) and MNs (blue circles) in the fiducial sample of Kepler transits.
The gray contours show the number density, which was smoothed using Gaussian kernel density estimation.
The black dashed line indicates the empirical position of the radius valley (see Eq.~\ref{es8}) for an illustrative host star of one solar mass.
Classification of each exoplanet used the actual observed mass of its host star\cite{methods}.
{\bf (B)} Mean orbital periods and eccentricities of MNs derived using the TDR method for each orbital period bin.
{\bf (C)} Same as panel (B), but for SEs.
All error bars indicate $1~\sigma$ uncertainties
}
\label{f1}
\end{figure}

\noindent We investigated the $P$--$e$ relation of small exoplanets (i.e., those smaller than four Earth radii) orbiting main-sequence stars of spectral types F, G, and K\cite{methods}.
We adopt a catalogue of transiting exoplanets observed using the Kepler space telescope\cite{methods,2018ApJS..235...38T,KDR25} and then split this sample into SEs and MNs (Fig.~\ref{f1}A) using the empirical position of the radius valley (see Eq.~\ref{es8}).

We statistically inferred the eccentricity distributions of these transiting exoplanets from their distributions of transit duration ratio (TDR)\cite{2008ApJ...678.1407F} as follows:
\begin{equation}
{\rm TDR} = \frac{T}{T_0} \sim \frac{\sqrt{(1 - b^2)(1 - e^2)}}{1 - e {\rm sin} \omega},
\label{eq1}
\end{equation}
where $T$ is the observed transit duration, $T_0$ is the expected transit duration for an exoplanet on an equivalent circular orbit (with the observed orbital period) transiting centrally across the stellar disk, $\omega$ is the argument of periapsis of the exoplanet's orbit (i.e., the angle from the orbit's ascending node to periapsis measured in the direction of the exoplanet's motion), and $b$ is the impact parameter of the transit. 
For an individual exoplanet, there are three unknown parameters in this equation, $e$, $\omega$ and $b$, so the eccentricity cannot be uniquely determined.
However, for a sufficiently large population, the distributions of $\omega$ and $b$ can be inferred from models\cite{2016PNAS..11311431X}, leaving only a single unknown: the eccentricity distribution. 
Therefore, the TDR distribution of a planetary population can be used to statistically reconstruct that population's underlying eccentricity distribution.
For a statistical population of transiting exoplanets, the values of $b$ approximately follows a uniform distribution between 0 and 1\cite{2016MNRAS.463.1323K}.
A population of transiting exoplanets on orbits with low eccentricities would thus have a TDR distribution concentrated close to TDR=1 (i.e., the limit of Eq.~\ref{eq1} as $e \to 0$), whereas those on orbits with higher eccentricities would have a more spread-out TDR distribution.

We applied this method to our fiducial sample of SEs and MNs by determining the TDR--$P$ distributions of each population.
We find that MNs with shorter orbital periods tended to have more spread out TDR values, whereas SEs showed the opposite trend (Fig.~\ref{fs1}).

We divided the SEs and MNs into nine orbital period bins and calculated the mean eccentricities of exoplanets in each using the TDR method\cite{2016PNAS..11311431X,2023AJ....165..125A,methods}.
For MNs (Fig.~\ref{f1}B), we find that $e$ decreases as $P$ increases, except for the first bin (with $P<4$~days).
We attribute this discrepant bin to a combination of its high uncertainty and a tidal damping effect.
The tidal forces between a star and a planet gradually dampen the planet's orbital eccentricity, a process known as tidal circularization.
The tidal circularization timescales of MNs at $P\approx4$~days are $<1$ billion years (much shorter than the typical age of host stars in our sample), which would produce a peak in the $P$--$e$ diagram (see Supplementary Text).
For SEs (Fig.~\ref{f1}C), we found that $e$ increase as a function of $P$.
 \end{multicols}

\begin{multicols}{2} 
\subsubsection*{Comparison with directly measured eccentricities}
\begin{figure}[!ht]
\centering
\includegraphics[width=1\textwidth]{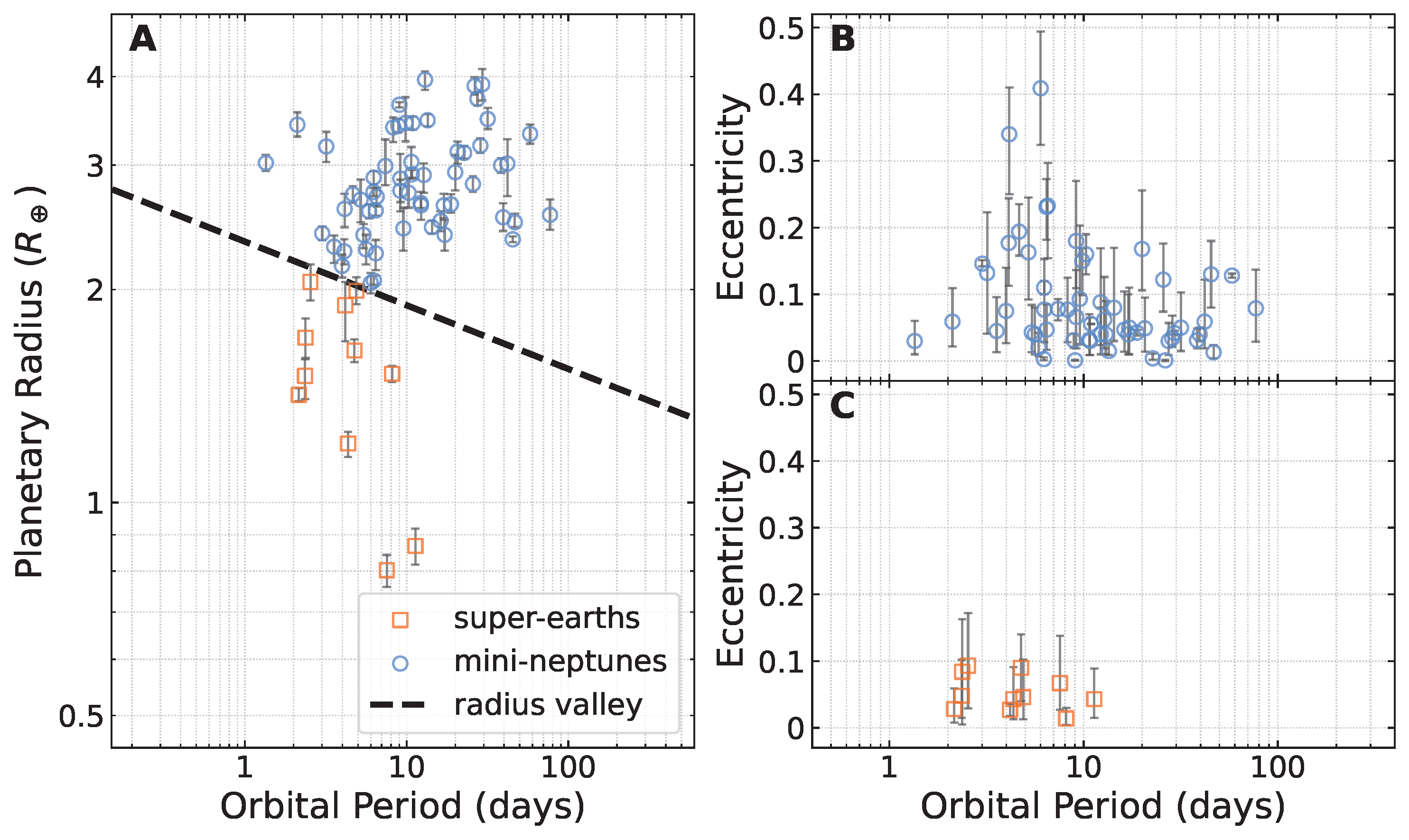}
\caption{
\textbf{Radii and eccentricities of the comparison NEA sample.}
({\bf A} to {\bf C}) Same as Fig.~\ref{f1} but for the comparison sample of individual eccentricity measurements reported in the NEA database\cite{NEA2024}.
All data points are individual measurements and have not been binned.
Error bars indicate the $1~\sigma$ uncertainties from previous studies as reported in the NEA database. 
}
\label{f2}
\end{figure}

\noindent As a consistency check, we compiled a sample of transiting small exoplanets with precise eccentricity measurements reported in the NASA Exoplanet Archive (NEA) database\cite{2013PASP..125..989A,NEA2024}. 
Unlike the fiducial Kepler sample, the NEA sample consists of exoplanets with individual eccentricity measurements reported by previous studies.
After application of our selection criteria\cite{methods}, there were 69 exoplanets in this comparison sample.
We split them into MNs and SEs using the same empirical radius valley (Fig.~\ref{f2}A).

MNs in the comparison sample follow a similar $P$--$e$ trend (Fig.~\ref{f2}B) as in the fiducial sample, with eccentricity generally decreasing with orbital period except for $P\lesssim4$~days.
For SEs, the comparison sample contains only 11 exoplanets, all with $P\lesssim11$~days and low eccentricities ($e\lesssim0.1$).
The resulting $P$--$e$ relation (Fig.~\ref{f2}C) is consistent with that of SEs in the fiducial sample (Fig.~\ref{f1}C), but the data are too limited to draw further conclusions.
We therefore exclude SEs in the comparison sample from our subsequent analyses.
\end{multicols}

\begin{multicols}{2}
\subsubsection*{Two distinct \emph{P--e} relations}
\noindent To quantify the $P$--$e$ relations of MNs and SEs, we fitted the $P$--$e$ distributions with a power-law model: $e = k\cdot(P/(10~{\rm days)})^c$, where $k$ (the amplitude) and $c$ (the power index) are the fitting parameters.
For the fiducial sample, we re-sampled the mean eccentricity of each orbital period bin by drawing $10^5$ trials from the posterior probability distribution output by the TDR analysis (Fig.~\ref{fs2}).
In each trial, we fitted the re-sampled $\bar{P}$--$\bar{e}$ distributions (where the bar symbols indicate the mean value in each bin) to obtain values for $k$ and $c$.
From the $10^5$ re-sampling processes, we fitted $k$ and $c$ $10^5$ times.
We report their resulting median values, with uncertainties corresponding to the 68.3\% confidence intervals.
To mitigate the effects of tidal damping, we excluded MNs with $P<4$~days (see the Supplementary Text).
The results were as follows:
\end{multicols}

\begin{equation}
\bar{e} =
\begin{cases}
0.165^{+0.019}_{-0.017}\cdot\left(\frac{\bar{P}}{10\;{\rm days}}\right)^{-0.45^{+0.13}_{-0.15}} {\rm ~for~MNs~} (R>R_{\rm valley}) \\
0.094^{+0.018}_{-0.018}\cdot\left(\frac{\bar{P}}{10\;{\rm days}}\right)^{0.37^{+0.34}_{-0.20}} {\rm ~~~for~SEs~} (R<R_{\rm valley})
\end{cases} 
,
\end{equation}
\begin{multicols}{2}
\noindent where $R$ and $R_{\rm  valley}$ are, respectively, the planet radius and the position of the radius valley (Eq.~\ref{es8}).
We apply a similar re-sampling and fitting analysis to MNs from the comparison sample\cite{methods} using the uncertainties on the individual eccentricity measurements reported in the NEA database\cite{NEA2024}.
The results of fitting comparison sample are shown in Fig.~\ref{f3}B.

Fig.~\ref{f3} compares the best-fitting power-law models with the data.
MNs in the fiducial sample (Fig.~\ref{f3}A) and comparison samples (Fig.~\ref{f3}B) followed a similar $P$--$e$ anti-correlation.
SEs follow an opposite trend (Fig.~\ref{f3}C).
The posterior probability distributions of the power-law indices for MNs and SEs (Fig.~\ref{f3}D) are separated, with minimal overlap.
The fraction of $c \geq 0$ for MNs is $9.6 \times 10^{-4}$ for the fiducial sample and $7.9 \times 10^{-3}$ for the comparison sample, corresponding to a $P$--$e$ anti-correlation ($c < 0$) at significance levels of $\sim3.3~\sigma$ and $\sim2.7~\sigma$, respectively.  

For the SEs in the fiducial sample, the fraction of $c \geq 0$ was 0.982, which indicates a $P$--$e$ positive correlation ($c > 0$) at a significance level of $\sim2.4~\sigma$.
Although this is only weak evidence of a positive correlation, the difference between SEs and MNs was more significant.
The posterior probability distributions of $c$ for SEs and MNs in the fiducial sample (Fig.~\ref{f3}D) overlapped by a fraction of $1.9 \times 10^{-4}$, corresponding to a significance level of $\sim 3.7~\sigma$.
We therefore conclude that SEs and MNs follow different $P$--$e$ relations.

To verify this conclusion, we investigated the effects on the derived $P$--$e$ relations from data binning\cite{methods}, discarding some data points and changing the position of the radius valley, our selection criteria for orbital period and signal-to-noise ratio, observational bias caused by the heterogeneity of the comparison sample, and the use of single-transit exoplanet systems in the fiducial sample (see the Supplementary Text). 
We found that none of these effects change our conclusions.

Previous studies of the $P$--$e$ distributions of small exoplanets\cite{2015ApJ...808..126V,2019AJ....157...61V,2020A&A...635A..37C,2024AJ....168..115B} did not report these relations.
We attribute this difference from our results to two main causes: previous work (i) studied systems with less precise individual eccentricity measurements (Fig.~\ref{fs21}) and (ii) did not separate SEs from MNs (see the Supplementary Text).

\subsubsection*{Comparison with theoretical models}
{\noindent \bf SEs}\\
Dynamical processes that occur during close encounters between planets are known as planet-planet scattering (PPS).
Theoretical models of PPS have predicted positive correlations between eccentricity and orbital period, as well as between eccentricity and planet mass\cite{2008ApJ...686..621F,2008ApJ...686..580C,2025ApJ...991L..49K}.

A simple model of PPS predicts a $P$--$e$ correlation with $c=1/3$ for a population of equal-mass planets (see the Supplementary Text). 
However, most small planets are observed in systems that have more massive outer planets\cite{2013ApJ...763...41C,2022MNRAS.514.3844C}.
We therefore expect $c>1/3$ for more realistic populations affected by PPS.
We adopted a minimum-mass extrasolar nebula (MMEN) model for small planets, which predicts that planetary mass increases with orbital period\cite{2020AJ....159..247D}.
Combining the MMEN model with the PPS theory, we derive a positive $P$--$e$ correlation with $c\approx0.42$ (see the Supplementary Text).
Fig.~\ref{f3}C shows this PPS prediction is consistent with the fiducial sample of SEs.
\end{multicols}

\begin{figure}[!ht]
\centering
\includegraphics[width=1\textwidth]{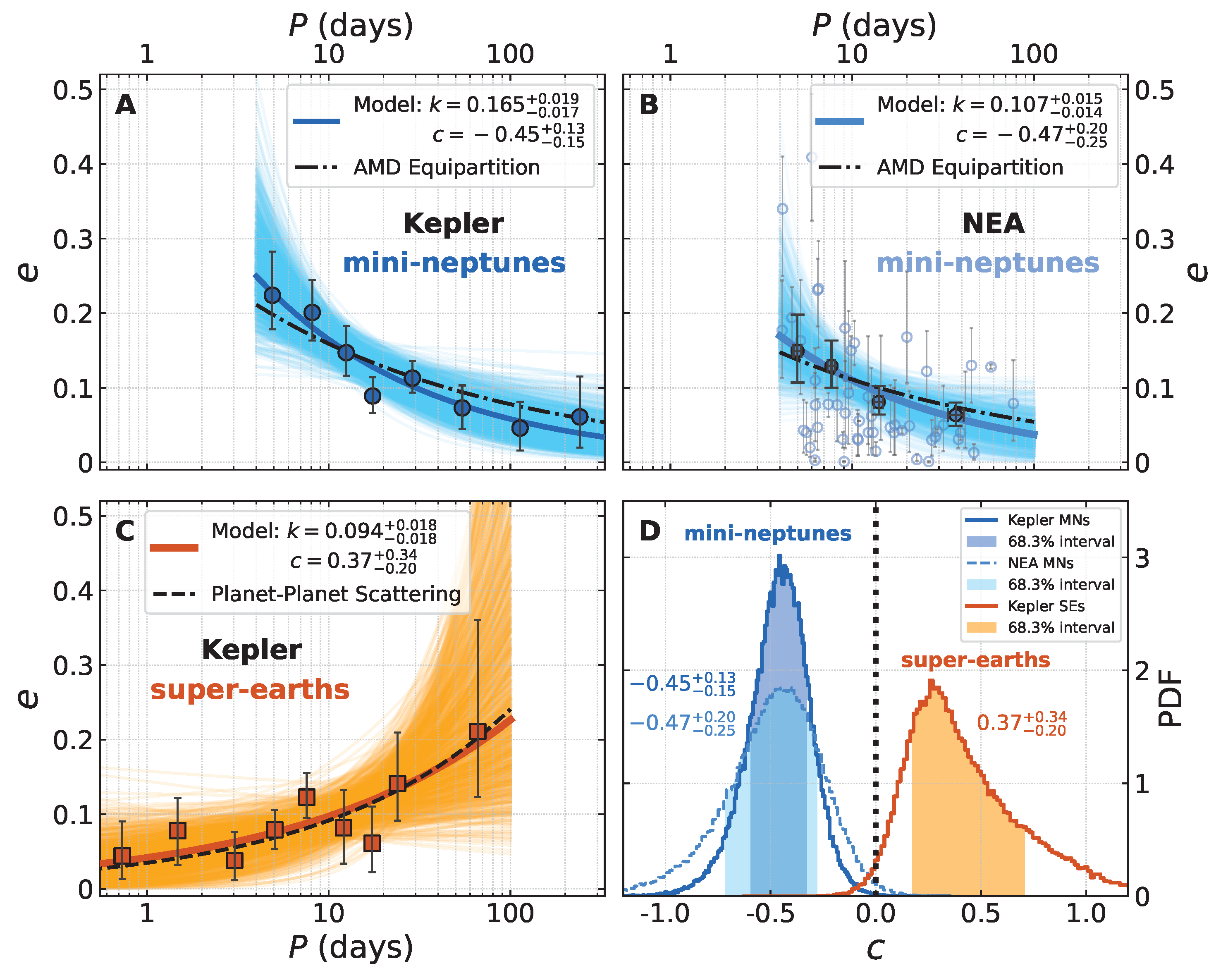}
\caption{
\textbf{Comparison of the period--eccentricity relations to theoretical models.}
{\bf (A)} Filled circles are mean eccentricities (with error bars indicating $1~\sigma$ uncertainties) and mean orbital periods for each orbital period bin in the fiducial Kepler sample.
The dark blue line is the best-fitting power-law model $e = k\cdot(P/{\rm 10~days})^c$, which has the parameters listed in the legend.
Light blue lines are 1000 equivalent models randomly drawn from the $10^5$ re-sampling processes.
The black dot-dashed line is a theoretical prediction for AMD equipartition (see Eq.~\ref{es22}).
{\bf (B)} Same as panel (A), but for MNs in the comparison NEA sample. 
Error bars on the mean orbital period are the 68.3\% confidence intervals of its posterior probability distributions\cite{methods}.
Open circles are individual measurements in the comparison NEA sample (as in Fig.~\ref{f2}B) but excluding those with $P<4$ days. 
{\bf (C)} Same as panel (A) but for SEs in the fiducial Kepler sample.
The black dashed line is a theoretical prediction for PPS (see Eq.~\ref{es17}).
{\bf (D)} Posterior probability distributions (histograms) of the fitted power-law index $c$ corresponding to panel (A) (solid dark blue), panel (B) (dashed light blue) and panel (C) (solid orange).
Colored shading indicates the corresponding 68.3\% confidence regions. 
}
\label{f3}
\end{figure}

\begin{multicols}{2}
Exoplanets with short orbital periods are expected to experience substantial tidal damping, which lowers their eccentricities.
This effect is also consistent with our measurement showing that the SEs on short orbital periods ($P \lesssim 6$~days) all exhibit low eccentricities ($\bar{e} \lesssim 0.1$) (Fig.~\ref{f1}C).
This could indicate that tidal damping also plays a role in establishing the $P$--$e$ positive correlation of SEs  (see the Supplementary Text). 
We were unable to distinguish between these two possibilities, so we conclude that the observed $P$--$e$ distribution of SEs is consistent with the predictions of PPS, tidal damping, or both mechanisms acting simultaneously.

{\noindent \bf MNs}\\
Long-term (secular) gravitational interactions between multiple planets in the same system are predicted to redistribute angular momentum.
This effect is expected to dominate in multi-planet systems with well-spaced orbits\cite{2011ApJ...735..109W,2017A&A...605A..72L}.
Theoretical models quantify this effect using the angular momentum deficit (AMD), the difference in the normal component of the angular momentum between the planet's orbit and a reference orbit (a coplanar circular orbit with the same semi-major axis)\cite{1997A&A...317L..75L,2017A&A...605A..72L}.
The total AMD in a planetary system is conserved, with any AMD exchange between planets acting towards equipartition\cite{2011ApJ...735..109W}.
Planets in a system that has reached AMD equipartition all have a similar amount of AMD, which requires their eccentricity to be anti-correlated with both the orbital period and planet mass\cite{2011ApJ...735..109W,2016ApJ...832...34M,2017A&A...605A..72L}.
For a population of equal-mass planets, full AMD equipartition would produce a $P$--$e$ anti-correlation with $c=-1/6$ (see the Supplementary Text).
As above, we expect a more realistic distribution of planet masses to produce $c<-1/6$.
By combining the observed period--mass relation with AMD equipartition theory, we derive a $P$--$e$ anti-correlation with $c\approx-0.31$ (see the Supplementary Text).
In Fig.~\ref{f3}, panel (A) and (B) show that this expectation is consistent with both samples of MNs.

The $P$--$e$ relation of MNs could also be affected by the host star’s metallicity (i.e., abundance of elements heavier than hydrogen and helium in a star).
Small planets with shorter orbital periods are observed to preferentially orbit stars with higher metallicity\cite{2016AJ....152..187M}, and small planets around high metallicity stars tend to have higher orbital eccentricities\cite{2023AJ....165..125A}.
We expect the combination of these two effects to cause some level of $P$--$e$ anti-correlation.
Using stellar metallicities measured by LAMOST (Large Sky Area Multi-Object Fiber Spectroscopic Telescope)\cite{2012RAA....12.1197C} and a Kepler-LAMOST small single-transit planet sample from previous work\cite{2023AJ....165..125A}, we perform a correction for the expected metallicity effect (see the Supplementary Text).
We found that the $P$--$e$ anti-correlation persisted, with a corrected $c=-0.39^{+0.14}_{-0.15}$ for MNs.
This value is closer to the prediction for AMD equipartition than the uncorrected measurement.

We also expect MNs with shorter orbital periods and higher eccentricities to be more strongly affected by tidal damping.
This would decrease the orbital eccentricities of MNs with short orbital periods (see the Supplementary Text), as we observed for $P\lesssim4$~days (Figs.~\ref{f1}B \& \ref{f2}B).
We therefore conclude that the observed $P$--$e$ distribution of MNs is consistent with the predictions of AMD equipartition, modified by tidal damping at short orbital periods.

\subsubsection*{Implications for exoplanet formation and evolution}
\noindent The consistency between our results for SEs and the predictions of PPS theory implies that violent processes such as giant impacts, collisions, and fragmentation dominate the orbital evolution of SEs.
These processes are known to influence planetary system architecture (i.e., the number of planets, their orbital eccentricities, and their orbital inclinations) and planet compositions\cite{2016ApJ...822...54D,2017AJ....154...27M}.
Compared with MNs, we expect collisions between SEs to produce more surviving large fragments because of their lower escape velocities and generally shorter orbital periods.
Observations have shown that giant impacts persist in at least some mature systems ($\gtrsim 1$ billion years old), although such events are rare at late epochs \cite{2015ApJ...805...77M}.
During these dynamic events, SEs could experience substantial atmospheric erosion, losing a large proportion of their volatile envelopes (such as a hydrogen and helium or steam atmosphere).
This erosion could drive SEs towards compositions similar to rocky planets with thin or absent atmospheres\cite{2015ApJ...812..164L,2015Icar..247...81S,2024ApJ...967...38Z}.

For MNs, the observed $P$--$e$ anti-correlation is consistent with the predictions of AMD equipartition theory, which implies that the evolution of MNs is dominated by secular interactions.
We expect secular interactions to dominate only in dynamically quiet systems, with minimal influence from PPS and other violent processes.
Therefore, MNs might have retained the majority of their volatile envelopes, leading to a lower bulk density than SEs.
For any MNs residing in a more chaotic environment, where PPS and giant impacts dominated their evolution, their resulting dynamical state and $P$--$e$ relation would be primarily influenced by PPS.
Such exoplanets could experience removal of their volatile envelopes and transition towards a more SE-like composition\cite{2015ApJ...812..164L,2015Icar..247...81S,2024ApJ...967...38Z}.    

Although the $P$--$e$ relation of SEs is not consistent with AMD equipartition, this does not preclude the presence of secular interactions in SE systems; it only indicates that they have not reached equilibrium.
We expect SEs to experience stronger tidal damping and greater influence from PPS than MNs. 
Tidal damping preferentially circularizes short-period orbits, whereas PPS increases eccentricities at longer periods.
This combination could produce a positive $P$--$e$ correlation for SEs. 

In summary, we found that SEs and MNs exhibit different $P$--$e$ relations.
This indicates that the two populations are in different dynamical states and thus might have experienced different dynamical evolutionary histories.
We suggest that SEs are more likely to have been affected by (or produced from MNs during) violent events, such as PPS and giant impacts.
By contrast, the orbits of surviving MNs might have been shaped by secular interactions driving them toward the equipartition of AMD.
\end{multicols}
~\\
\begin{multicols}{2}

%
\bibliography{references} 
\bibliographystyle{references}

%
%
%
%
%
%
\end{multicols}
~\\

\begin{multicols}{2}
\section*{Acknowledgments}
\noindent We thank the editors and referees for their insightful comments and suggestions.
We also thank Yanqin Wu, Hui-Gen Liu, Jia-Yi Yang, Di-Chang Chen, Yuan-Zhe Dai, and An-Dong Chen for helpful discussions.
This work has made use of the NEA, which is operated by the California Institute of Technology, under contract with the National Aeronautics and Space Administration under the Exoplanet Exploration Program.
This work made use of the data from LAMOST (also known as the Guoshoujing Telescope, https://lssf.cas.cn/en/facilities/sa/lamost).
LAMOST is a Chinese national mega-science facility, operated by National Astronomical Observatories, Chinese Academy of Sciences.
\paragraph*{Funding:}
J.-W.X  was supported by the National Key R\&D Program of China (grant number 2024YFA1611803) and the National Natural Science Foundation of China (grant number 12273011). 
J.-L.Z was supported by the National Key R\&D Program of China (grant number 2024YFA1611800).
\paragraph*{Author contributions:}
J.-W.X. conceived the project and designed the research.
J.-W.X. initially identified the anti-correlation between $P$ and $e$ of small planets from the NEA.
K.-T.S. and J.-W.X. performed tests and analysis of this relation for MNs using NEA data.
D.-S.A. and J.-W.X. validated the $P$--$e$ anti-correlation for MNs and identified the correlation for SEs by analyzing the Kepler data.
K.-T.S. and J.-W.X. developed the theoretical models and interpreted the observed $P$--$e$ relation.
K.-T.S., D.-S.A., and J.-W.X. drafted the manuscript.
J.-L.Z and F.D. discussed the results and revised the manuscript.
\paragraph*{Competing interests:}
There are no competing interests to declare.
\paragraph*{Data and materials availability:}
Our input data were obtained from Kepler data release 25\cite{KDR25} and the NEA database\cite{NEA2024}.
After application of our selection cuts and classification, our fiducial and comparison samples have been provided as Data S1 and Data S2\cite{zenodo}.
Our software codes for TDR analysis, bootstrap re-sampling, and generating simulated transiting exoplanet systems are archived at Zenodo\cite{zenodo}.
No physical materials were generated in this work.

\subsection*{Supplementary Materials}
\noindent Materials and Methods\\
Supplementary Text\\
Figures S1 to S21\\ 
Tables S1 to S2\\
References \textit{(47-\arabic{enumiv})}\\ 

\end{multicols}
\newpage

\renewcommand{\thefigure}{S\arabic{figure}}
\renewcommand{\thetable}{S\arabic{table}}
\renewcommand{\theequation}{S\arabic{equation}}
\renewcommand{\thepage}{S\arabic{page}}
\setcounter{figure}{0}
\setcounter{table}{0}
\setcounter{equation}{0}
\setcounter{page}{1} 

\begin{center}
\section*{Supplementary Materials for\\ \scititle}

Ke-Ting Shin (辛科霆)$^{\dagger}$,
Dong-Sheng An (安东升)$^{\dagger}$,
Ji-Wei Xie (谢基伟)$^{\ast}$,\\
Ji-Lin Zhou (周济林),
Fei Dai (戴飞)\\
\small$^\dagger$These authors contributed equally to this work.\\
\small$^\ast$Corresponding author. Email: jwxie@nju.edu.cn
\end{center}

\subsubsection*{This PDF file includes:}
\noindent Materials and Methods\\
Supplementary Text\\
Figures S1 to S21\\
Tables S1 to S2\\
\newpage

{\noindent \bf \fontsize{13pt}{15pt} \selectfont Materials and Methods}\\
\hyperlink{s1}{\bf \emph{P--e} relation from TDR analysis}

\hyperlink{s1.1}{The fiducial Kepler sample} (Table S1)

\hyperlink{s1.2}{TDR method} (Equations S1 to S7)

\hyperlink{s1.3}{Results of TDR analysis} (Figures S1 to S3, equation S8)

\hyperlink{s1.4}{Effect of Binning} (Figures S4 to S7)\\
\hyperlink{s2}{\bf \emph{P--e} relation from individual measurement}

\hyperlink{s2.1}{The comparison NEA sample} (Table S2)

\hyperlink{s2.2}{Effect of eccentricity uncertainty cut} (Figures S8 to S9, equation S9)\\
{\noindent \bf \fontsize{13pt}{15pt} \selectfont Supplementary Text}\\
\hyperlink{s3}{\bf Theoretical \emph{P--e} relation}

\hyperlink{s3.1}{Planet-planet scattering} (Equations S10 to S17)

\hyperlink{s3.2}{Angular momentum deficit equipartition} (Figure S10, equations S18 to S22)

\hyperlink{s3.3}{Effect of Tidal Damping} (Equations S23 to S26)\\
\hyperlink{s4}{\bf Data point dropout test} (Figures S11 to S12)\\
\hyperlink{s5}{\bf Effect of changing radius boundary} (Figures S13 to S14)\\
\hyperlink{s6}{\bf Effect of orbital period cut} (Figure S15)\\
\hyperlink{s7}{\bf Effect of SNR cut} (Figures S16 to S17)\\
\hyperlink{s8}{\bf Effect of observational bias}

\hyperlink{s8.1}{TTV systems and MMR systems} (Figure S18, equation S27)

\hyperlink{s8.2}{Effect of using single-transit systems} (Figure S19, equations S28 to S29)\\
\hyperlink{s9}{\bf Effect of stellar metallicity} (Figure S20, equations S30-31)\\
\hyperlink{s10}{\bf Previous studies} (Figure S21)
\clearpage

\begin{multicols}{2}
\section*{Materials and Methods}
\subsection*{\hypertarget{s1}{\emph{P--e} relation from TDR analysis}}
\subsubsection*{\hypertarget{s1.1}{The fiducial Kepler sample}}
\noindent Our fiducial Kepler sample is built from Kepler data release 25 (Kepler DR25)\cite{2018ApJS..235...38T,KDR25}, which includes 8054 Kepler objects of interest (KOIs) in 6923 systems.
Table~\ref{ts1} provides an overview of the selection cuts we apply, which are described in detail below.

We first exclude all false positives flagged in the koi\_disposition column of Kepler DR25 catalog (criterion 1), then cross-match the remaining KOIs with a stellar parameter catalog\cite{2020AJ....159..280B}(criterion 2).
We set a maximum threshold of 1.2 on the renormalized unit weight error (RUWE) reported in Gaia data release 2\cite{2016A&A...595A...1G,2018A&A...616A...1G}, chosen to filter out potential binary stars (criterion 3).
We use Gaia data release 2 for consistency with the stellar parameters catalog\cite{2020AJ....159..280B}.
We have checked that using Gaia data release 3\cite{2023A&A...674A...1G} has a negligible effect on the results.
We also set an lower limit of 0.99 on the goodness of fit (GOF) parameter in the stellar parameter catalog to ensure the robustness of stellar parameters.
We then apply stellar parameter cuts of $\log_{10}(g\cdot{\rm s}^2\cdot{\rm cm}^{-1})>4$ on the surface gravity $g$, to select main sequence stars (criterion 5) and effective temperatures $4700{\rm K} < T_{\rm eff} < 6500{\rm K}$ (criterion 6) to focus on FGK-type stars.
This ensures that all exoplanets in our sample have similar host stars\cite{2017AJ....154..109F,2018ApJ...866...99B}.
After applying all the above criteria, we are left with 2651 exoplanets in 1978 systems.
\end{multicols}

\begin{table}[!ht]
\centering
\caption{
\textbf{Selection cuts for the fiducial Kepler sample.}
The first column is the sample selection criterion we add at each step to select systems from the Kepler DR25 dataset, and the second column shows the number of exoplanets remaining in the sample after applying the corresponding criterion.
In step 11, we also remove exoplanets without transit duration uncertainties.
}
\begin{tabular}{lc}
\\
\hline
\multicolumn{1}{c}{Criterion} & Number of exoplanets \\
\hline
0. Kepler DR25   & 8054 \\
1. Not false positive  & 4034 \\
2. Stellar properties cross-match\cite{2020AJ....159..280B} & 3826\\
3. Gaia RUWE $<1.2$ & 	3413 \\
4. GOF $>0.99$ & 	3401\\
5. $\log_{10}(g\cdot{\rm s}^2\cdot{\rm cm}^{-1})>4$  & 3123 \\
6. $4700~{\rm K} < T_{\rm eff} < 6500~{\rm K}$  & 2651 \\
7. SNR $>7.1$ & 2626 \\
8. Relative Error of $r \equiv \frac{R}{R_{\star}} < 0.3$  & 2552 \\
9. Single-transit systems & 1463 \\
10. Disposition Score $>0.9$ & 1270 \\
11. TDR $<1.5$ & 1266 \\
12. $R<4~R_\oplus$ & 1104 \\
\hline
\end{tabular}
\label{ts1}
\end{table}

\begin{multicols}{2}
We then apply two further criteria to refine the sample: a minimum transit signal-to-noise ratio (SNR, the koi\_model\_snr column in Kepler DR25 catalog) of 7.1 (criterion 7)\cite{2011ApJ...736...19B}, and a maximum relative error of 0.3 (criterion 8) in the radius ratio ($r \equiv R/R_{\star}$, where $R$ and $R_\star$ are the radius of exoplanets and host stars, respectively)\cite{2019AJ....157..198M}.
The SNR cut was chosen to select KOIs with sufficiently well-measured planetary parameters (e.g. transit duration) and to exclude remaining false positives.
We test this choice below (see Supplementary Text).
After theses refinements, there are 1463 single-transit systems (containing 1463 exoplanets) and 441 multi-transit systems (containing 1089 exoplanets) in the sample.

We chose to utilize only single-transit systems (criterion 9) to derive the orbital $P$--$e$ relation of SEs and MNs.
This is due to a limitation of our method.
To derive the eccentricity of multiple-transit exoplanets with their TDR, we would need to account for the constraints from the mutual inclination of the exoplanets in each system as a whole\cite{2016PNAS..11311431X}.
However, to derive the $P$--$e$ relation of multiple-transit exoplanets, we have to divide the exoplanets into different data bins according to their orbital period.
This would be a contradiction, so we do not consider multi-transit systems any further.
However we expect the $P$--$e$ relation for single transit systems to be representative of all systems (see Supplementary Text).

We only include single-transit exoplanets with disposition scores (the koi\_score column in Kepler DR25 catalog) larger than 0.9 (criterion 10)\cite{2018ApJS..235...38T}, and exclude outliers with TDR > 1.5 (criterion 11).
To focus on small plants, we require exoplanet radii to be smaller than 4 Earth radii (criterion 12)\cite{2012ApJS..201...15H,2013ApJ...770...69P}.
We examine the effect of including larger exoplanets below (see Supplementary Text).

Our final fiducial Kepler sample contains 1104 exoplanets in 1104 systems (Fig.~\ref{f1}A).
This fiducial samples are provided as data S1\cite{zenodo}.

\subsubsection*{\hypertarget{s1.2}{TDR method}}
\noindent We estimate the eccentricity distribution of a group of transiting exoplanets from their TDR distribution\cite{2008ApJ...678.1407F}.
TDR is defined as the ratio of the planetary transit duration $T$ to a reference duration $T_0$ (Eq.~\ref{eq1}), which is the transit duration that an exoplanet would have if its orbit was circular and the transit impact parameter was zero (assuming $m \ll M_{\star}$, where $m$ and $M_{\star}$ are the mass of the exoplanet and the mass of the host star, respectively):
\begin{equation}
\label{es1}
T_0 = \frac{(R+R_\star) P}{\pi a} = \left(\frac{3}{{\rm G}\pi^2}\right)^{\frac{1}{3}} \left( \frac{P}{\rho_\star} \right)^{\frac{1}{3}}(1 + r),
\end{equation}
where $a$, $\rho_{\star}$ and G are the planetary orbital semi-major axis, the stellar density, and the universal gravitational constant, respectively.

We perform forward modeling to generate simulated TDR (${\rm TDR}_{\rm sim}$) following previous work\cite{2016PNAS..11311431X,2023AJ....165..125A}.
Then we apply the maximum likelihood estimation approach to fit ${\rm TDR}_{\rm sim}$ to the observed TDR (${\rm TDR}_{\rm obs}$).
The likelihood function is:
\end{multicols}

\begin{equation}
\label{es2}
\mathcal{L}({\rm TDR}_{\rm obs} | \bar{e}) = \int {\rm PDF}({\rm TDR}_{\rm sim} | \bar{e}) \cdot {\rm exp} \left[ -\frac{({\rm TDR}_{\rm sim} - {\rm TDR}_{\rm obs})^2}{2 \sigma_{{\rm TDR}}^2} \right] {\rm d} {\rm TDR}_{\rm sim},
\end{equation}

\begin{multicols}{2}
\noindent where $\bar{e}$ is the mean orbital eccentricity of a group of exoplanets, $\sigma_{\rm TDR}$ is the uncertainty of ${\rm TDR}_{\rm obs}$ and ${\rm PDF}({\rm TDR}_{\rm sim} | \bar{e})$ is the probability density function (PDF) of the simulated TDR (with mean eccentricity of $\bar{e}$).
By maximizing Eq.~\ref{es2}, we obtain the best fitting value of $\bar{e}$.
\end{multicols}
\begin{multicols}{2}
To estimate ${\rm PDF}({\rm TDR}_{\rm sim} | \bar{e})$ in Eq.~\ref{es2}, we randomly assign $e$, $\omega$ and $I_{\rm p}$ (the planetary inclination relative to the observer) for each exoplanet in our sample to obtain the distribution of ${\rm TDR}_{\rm sim}$.
Specifically, $e$ is drawn from a Rayleigh distribution, $\omega$ is drawn from a uniform distribution between 0 and $2\pi$, and $I_{\rm p}$ is assigned by assuming ${\rm cos}I_{\rm p}$ follows a uniform distribution between -1 and 1.
After the assignment, the transit impact parameter $b$ of each exoplanet is calculated via (assuming $m \ll M_{\star}$):
\end{multicols}

\begin{equation}
\label{es3}
b = \frac{a {\rm cos}I_{\rm p}}{R_\star}  \frac{1-e^2}{1-e {\rm sin}\omega} = \left(\frac{{\rm G}}{3\pi}\right)^{\frac{1}{3}} P^{\frac{2}{3}} \rho_\star^{\frac{1}{3}} \frac{1-e^2}{1-e {\rm sin}\omega} {\rm cos}I_{\rm p},
\end{equation}

\begin{multicols}{2}
\noindent where $\omega$ is the argument of periapsis of the exoplanet's orbit, differing by $\pi$ radians from the argument of periapsis of the star's orbit due to the exoplanet ($\omega_\star = \omega+\pi$).
The modeled transit duration of each exoplanet is calculated via:
\begin{equation}
\label{es4}
T = \frac{P}{\pi} \frac{(1 - e^2)^\frac{3}{2}}{(1 - e {\rm sin} \omega)^2} {\rm arcsin} \left[ \frac{\sqrt{(1 + r)^2 - b^2}}{b \tan I_{\rm p}} \right],
\end{equation}
To simulate an observation, we consider the observational uncertainties of $T_0$ and $T$.
We obtain the simulated transit duration by assuming it has the same relative uncertainty as the observation:
\begin{equation}
\label{es5}
T_{\rm sim} = T + R_{\rm N1} \sigma_{T_{\rm obs}} \frac{T}{T_{\rm obs}},
\end{equation}
where $\sigma_{T_{\rm obs}}$ is the uncertainty of $T_{\rm obs}$ and $R_{\rm N1}$ is a random variable drawn from a normal distribution with median of 0 and standard deviation of 1.
Similarly, we obtain simulated reference transit duration by assuming it has the same uncertainty as the observation:
\begin{equation}
\label{es6}
T_{0\rm sim} = T_0 + R_{\rm N2} \sigma_{T_0},
\end{equation}
where $\sigma_{T_0}$ is the uncertainty of $T_0$ and $R_{\rm N2}$ is a random variable drawn from a normal distribution with median of 0 and standard deviation of 1.
Following previous work\cite{2014ApJ...790..146F}, we assign a simulated SNR for each $T_{\rm sim}$:
\begin{equation}
\label{es7}
{\rm SNR}_{\rm sim} = {\rm SNR}_{\rm obs} \sqrt{\frac{T_{\rm sim}}{T_{\rm obs}}},
\end{equation}
where ${\rm SNR}_{\rm sim}$ and ${\rm SNR}_{\rm obs}$ are the simulated SNR and observed SNR, respectively.
If the exoplanet with these assigned parameters does not transit ($|b|>1+r$), crashes into the star ($a(1-e)<R_{\star}+R$) or with ${\rm SNR}_{\rm sim}<7.1$, we discard it and repeat the assignment above.
Otherwise, we continue to calculate the simulated TDR (${\rm TDR}_{\rm sim}=T_{\rm sim}/T_{0\rm sim}$).
These steps are repeated until at least 300 ${\rm TDR}_{\rm sim}$ are assembled for each exoplanet.
We then calculate the mean eccentricity ($\bar{e}$) of these exoplanets and use the Gaussian kernel density estimation (KDE) to estimate ${\rm PDF}({\rm TDR}_{\rm sim})$.
We vary the mean value of the Rayleigh distribution to obtain ${\rm PDF}({\rm TDR}_{\rm sim} | \bar{e})$.

\subsubsection*{\hypertarget{s1.3}{Results of TDR analysis}}
\begin{figure}[!ht]
\centering
\includegraphics[width=1\textwidth]{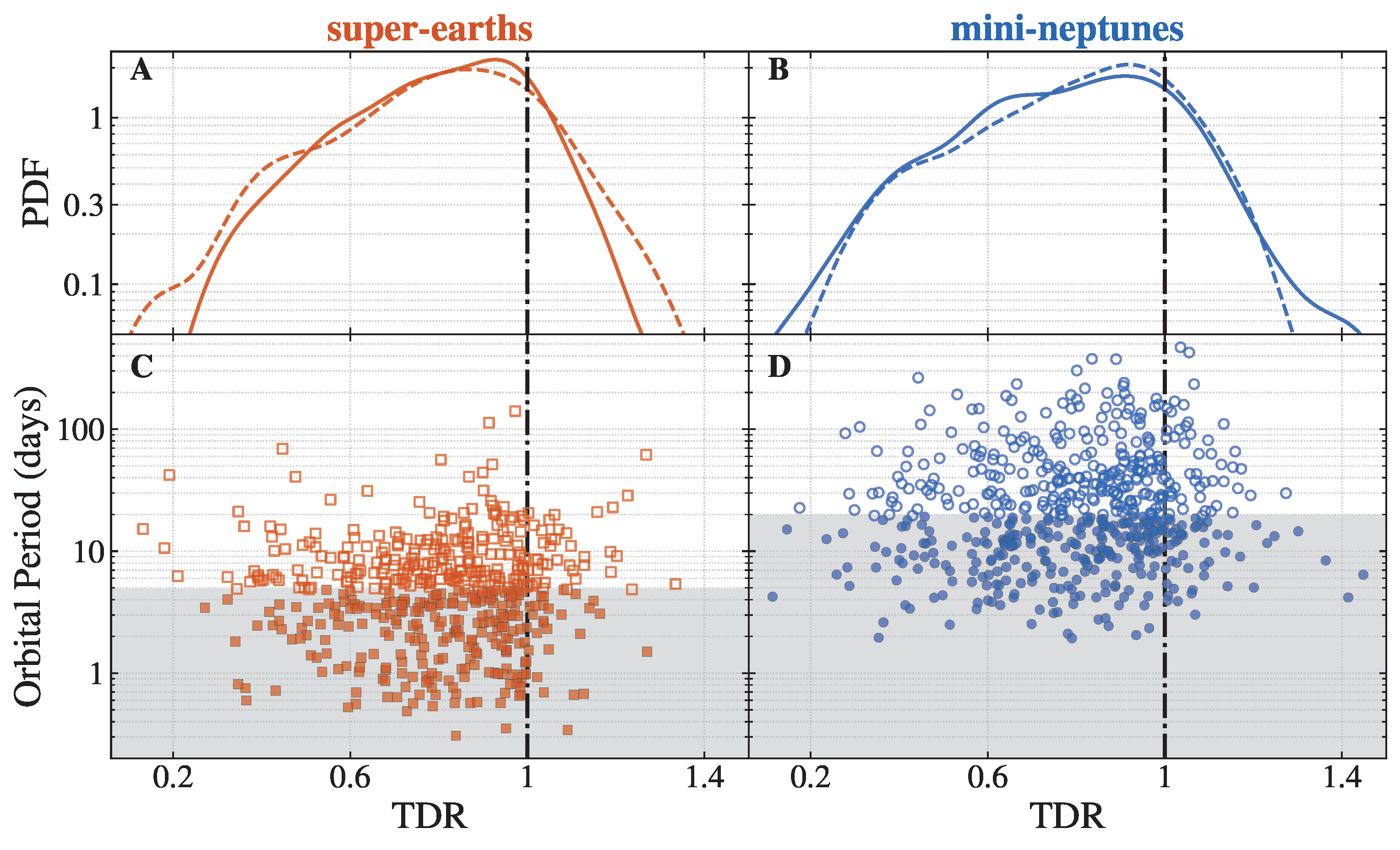}
\caption{
\textbf{The distribution of TDR and its dependence on the orbital period.}
SEs (orange lines and squares) and MNs (blue lines and circles) in the fiducial sample are divided into two sub-samples (with almost even size) according to their orbital periods.
{\bf (A)} The PDF of SEs' TDR (smoothed by Gaussian KDE).
The orange solid (dashed) lines are the PDF of SEs in the sub-samples with shorter (longer) orbital periods.
{\bf (B)} Same as panel (A), but for MNs in the fiducial sample.
{\bf (C)} The TDR--period distribution of SEs.
The filled (open) orange squares are the exoplanets in the sub-samples with shorter (longer) orbital periods.
The PDF of TDR of long-period SEs (open orange squares) is shown by the dashed orange line in panel (A), and the PDF of TDR of short-period SEs (filled orange squares) is shown by the solid orange line in panel (A).
The shaded region indicates the range of orbital periods for SEs in the short-period subsample (filled orange squares).
{\bf (D)} Same as panel (C), but for MNs in the fiducial sample.
}
\label{fs1}
\end{figure}

\noindent We adopt an empirical determination of the position of the radius valley\cite{2018MNRAS.479.4786V,2020AJ....160..108B,2021ARA&A..59..291Z}:
\begin{equation}
\label{es8}
R_{\rm valley} = 1.9~R_\oplus \; \left(\frac{P}{10\;{\rm days}}\right)^{-0.09} \left(\frac{M_{\star}}{M_\odot}\right)^{0.26},
\end{equation}
where $R_\oplus$ is the radius of Earth.
We use Eq.~\ref{es8} to divide our fiducial Kepler sample into two populations: SEs, with $R<R_{\rm valley}$ and MNs, with $R_{\rm valley}<R<4~R_\oplus$.
As a demonstration, we divide both SEs and MNs into two further sub-samples (with almost even size) according to their orbital periods.
Their TDR--$P$ distributions (Fig.~\ref{fs1}) show that SEs with shorter orbital periods tend to have narrower TDR distribution (TDR distributes closer to one), while MNs with shorter orbital periods tend to have a more dispersive TDR distribution (TDR distributes further from one).
This implies that SEs with shorter periods have lower eccentricities, while MNs with shorter periods have higher eccentricities.
\end{multicols}

\begin{multicols}{2}
To investigate the $P$--$e$ relations, we divide both SEs and MNs into nine bins according to their orbital periods.
The bin boundaries are (1, 2, 4, 6, 10, 15, 20, 40)~days for SEs and (4, 6, 10, 15, 20, 40, 80, 160)~days for MNs.
Then we apply the TDR method described above to derive the mean eccentricities of exoplanets in each bin.
To obtain a smoothed likelihood function of $\bar{e}$, we apply a least squares polynomial fit to its relative likelihoods.
The observed TDR distributions, ${\rm PDF}({\rm TDR}_{\rm sim} | \bar{e})$, and the likelihood functions of $\bar{e}$ for each bin are shown in Fig.~\ref{fs2} and Fig.~\ref{fs3}.
\end{multicols}
\clearpage

\begin{multicols}{2} 
\noindent
\includegraphics[height=1\textheight]{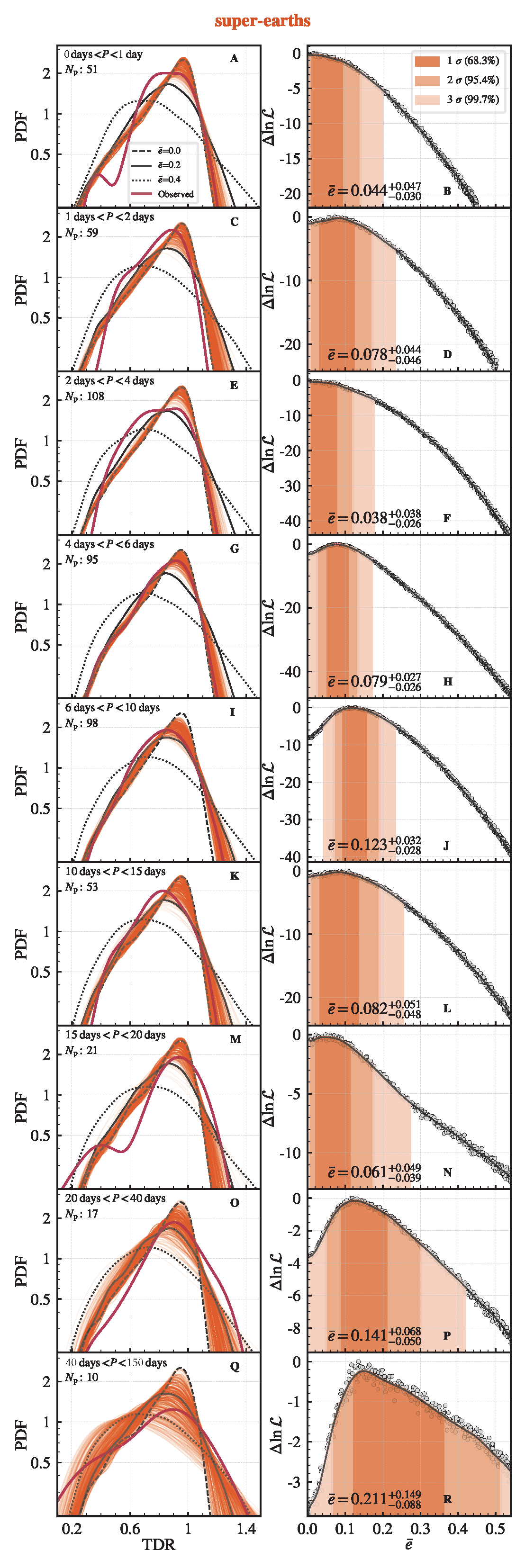}
\captionof{figure}{
\textbf{TDR fitting results of the fiducial Kepler sample.}
The first column shows the PDF of TDR from each bin (smoothed by Gaussian KDE).
Red lines are the distributions of observed TDR (${\rm TDR}_{\rm obs}$).
Orange lines are the ${\rm PDF}({\rm TDR}_{\rm sim} | \bar{e})$ with 1000 $\bar{e}$ randomly drawn from its posterior probability distribution.
Dashed, solid and dotted lines are the ${\rm PDF}({\rm TDR}_{\rm sim} | \bar{e})$ with $\bar{e}=0.0,\;0.2,\;0.4$.
The sizes (numbers of exoplanets) and orbital period boundaries of each bin are labeled in each panel.
The second column shows the relative likelihoods of $\bar{e}$.
Circles show the relative likelihoods calculated using Eq.~\ref{es2} and solid lines are the likelihood functions derived from polynomial interpolations.
Dark, medium and light colored shading indicates the 68.3\%, 95.4\%, and 99.7\% confidence intervals.
The median values of $\bar{e}$ with $1~\sigma$ uncertainties (68.3\% confidence intervals) are labeled in each panel.
}
\label{fs2}
\end{multicols}
\clearpage

\begin{multicols}{2} 
\noindent
\includegraphics[height=1\textheight]{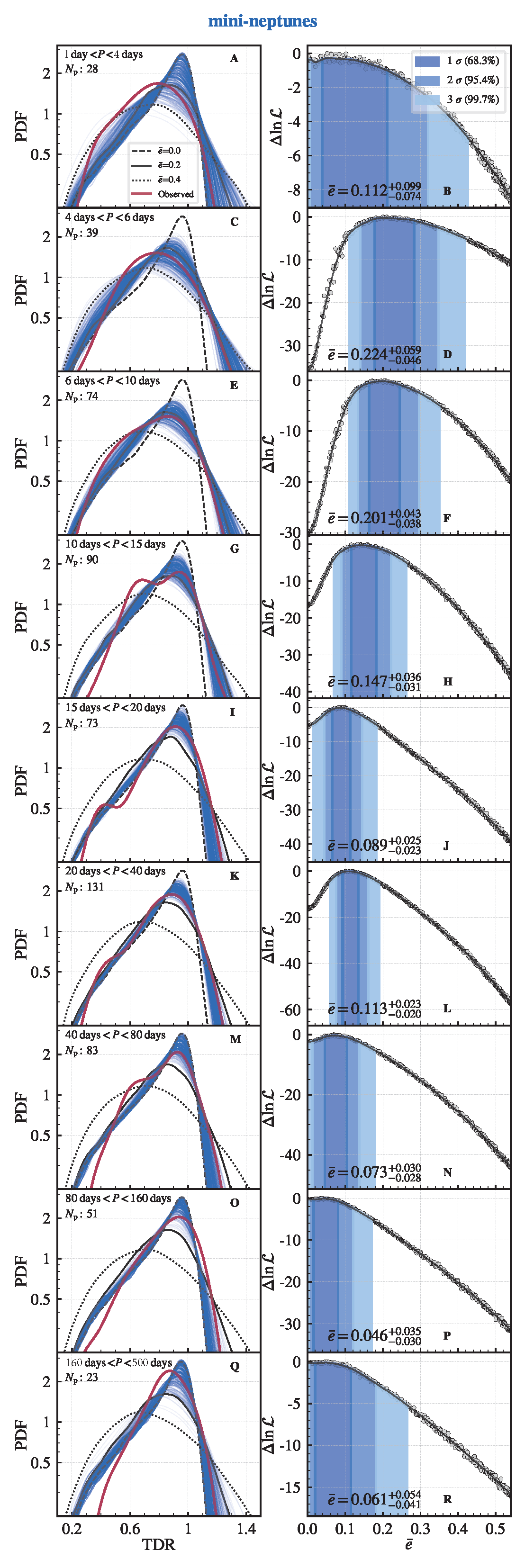}
\captionof{figure}{
\textbf{Same as Fig.~\ref{fs2}, bur for MNs in the fiducial Kepler sample.}
}
\label{fs3}
\end{multicols}
\clearpage

\begin{multicols}{2}
\subsubsection*{\hypertarget{s1.4}{Effect of binning}}
\begin{figure}[!ht]
\centering
\includegraphics[width=0.9\textwidth]{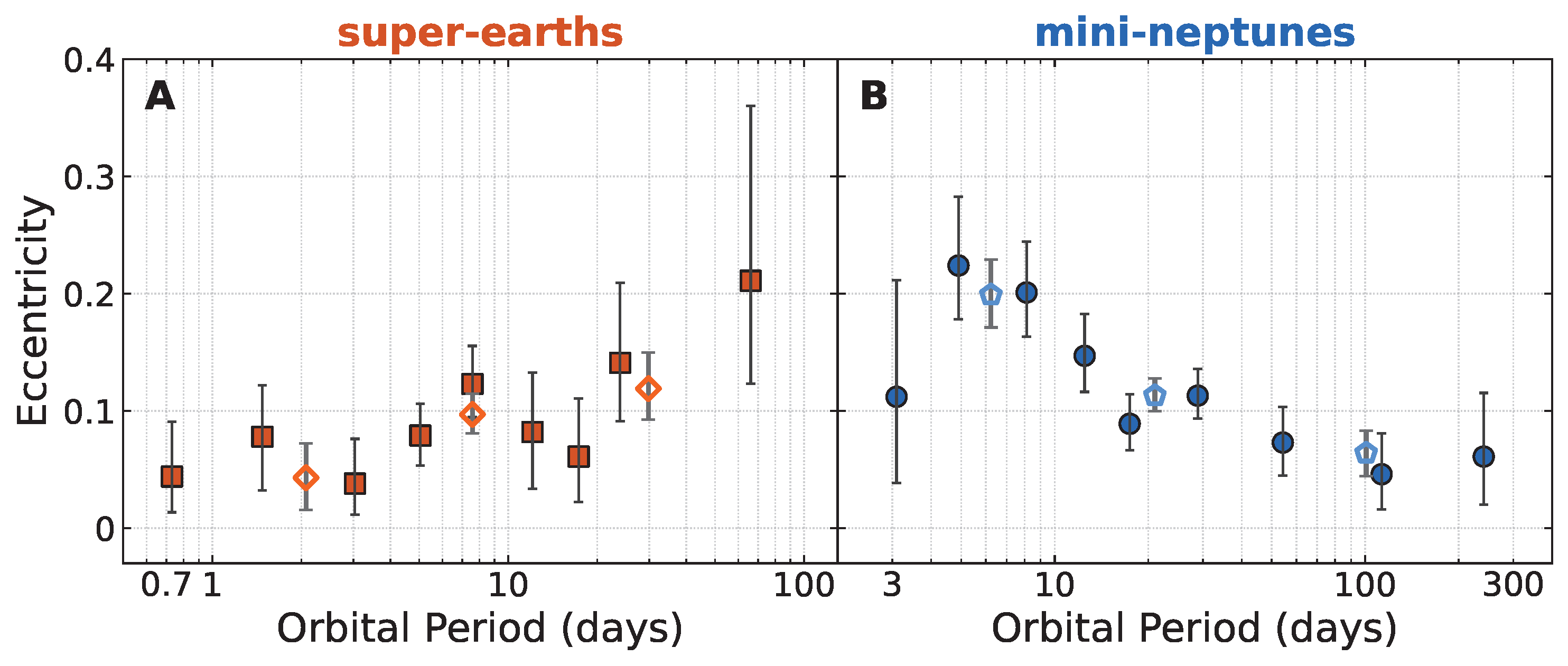}
\caption{
\textbf{Comparison between results using three and nine bins.}
Same as Fig.~\ref{f1}B\&C, but open diamonds (SEs) and open pentagons (MNs) show the results from three orbital period bins.
}
\label{fs6}
\end{figure}

\noindent To test whether the number (size) of data bins affects the derived $P$--$e$ relations, we re-analyzed the samples using three data bins, with bin boundaries (4, 15)~days for SEs and (10, 40)~days for MNs.
In this case, all data bins contain more than 40 exoplanets.
The observed TDR distributions, the ${\rm PDF}({\rm TDR}_{\rm sim} | \bar{e})$, and the likelihood functions of $\bar{e}$ for each bin are shown in Fig.~\ref{fs4} and Fig.~\ref{fs5}.
These larger bins result in smaller uncertainties.
The $P$--$e$ distributions from three bins remain consistent to those from nine bins: the SEs with shorter periods show lower eccentricities, while the MNs with shorter periods show higher eccentricities (Fig.~\ref{fs6}).

\begin{figure}[!ht]
\centering
\includegraphics[width=0.95\textwidth]{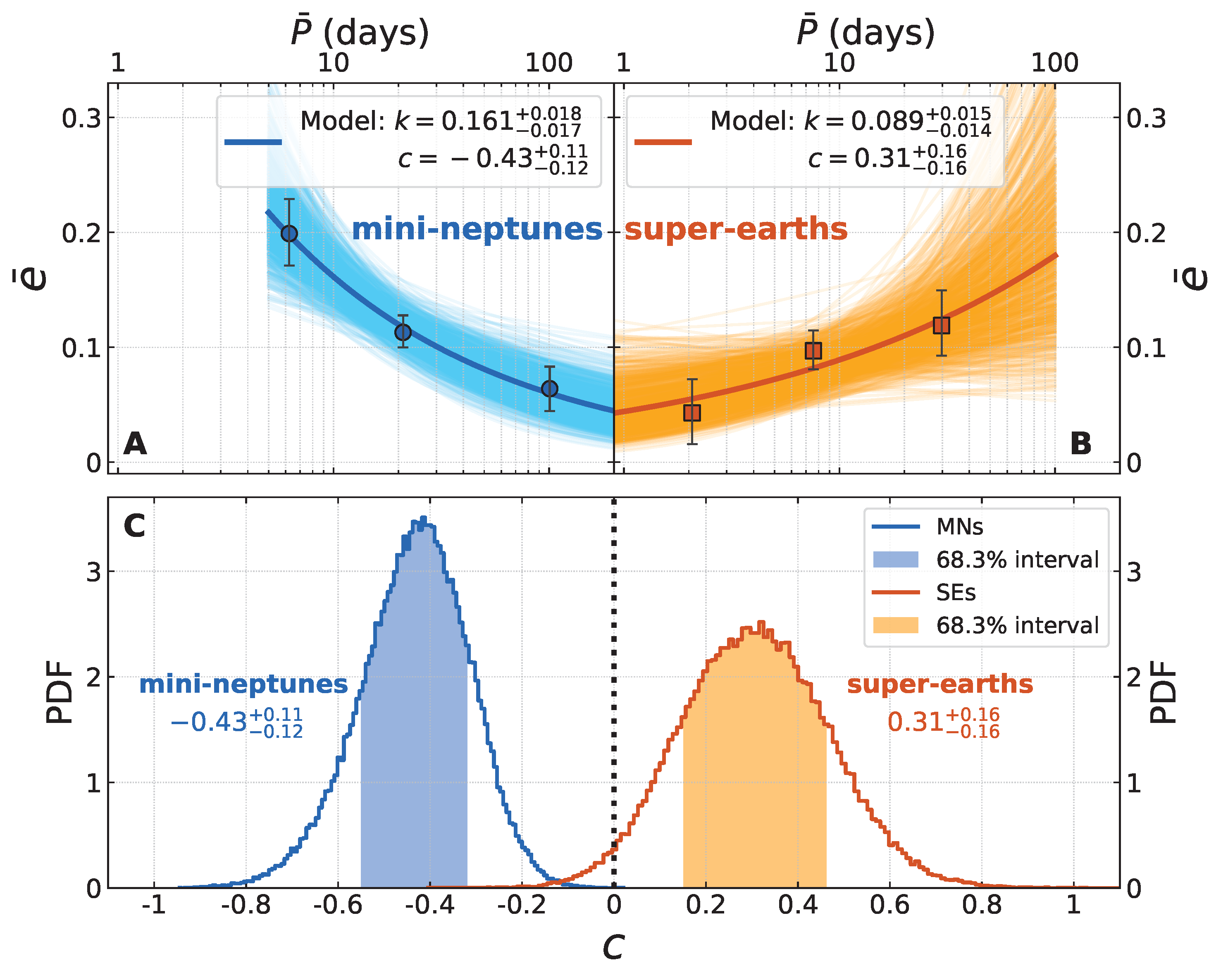}
\caption{
\textbf{Same as Fig.~\ref{f3}A, C\&D, but for the analysis in three bins.}
}
\label{fs7}
\end{figure}

We perform the same fitting procedures to derive the $P$--$e$ relations from the three bins' results.
Fig.~\ref{fs7} shows the resulting $P$--$e$ relations of SEs and MNs are consistent with those presented in Fig.~\ref{f3}.
Therefore, we conclude that the number of the data bins does not substantially affect our derived $P$--$e$ relations.
\end{multicols}

\begin{multicols}{2} 
\noindent
\includegraphics[width=0.5\textwidth]{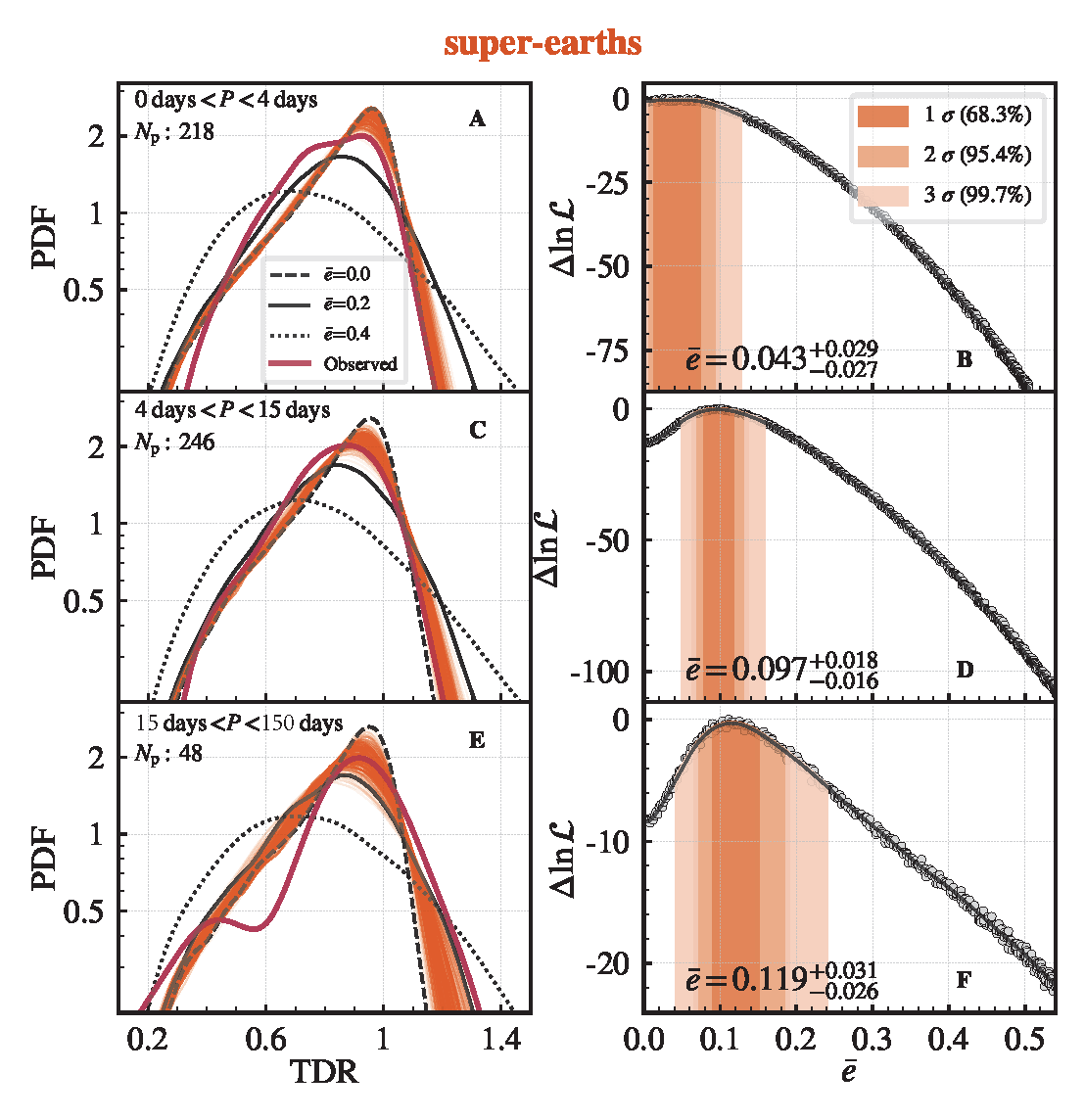}
\captionof{figure}{
\textbf{Same as Fig.~\ref{fs2}, bur for SEs in three orbital period bins.}
}
\label{fs4}

\noindent
\includegraphics[width=0.5\textwidth]{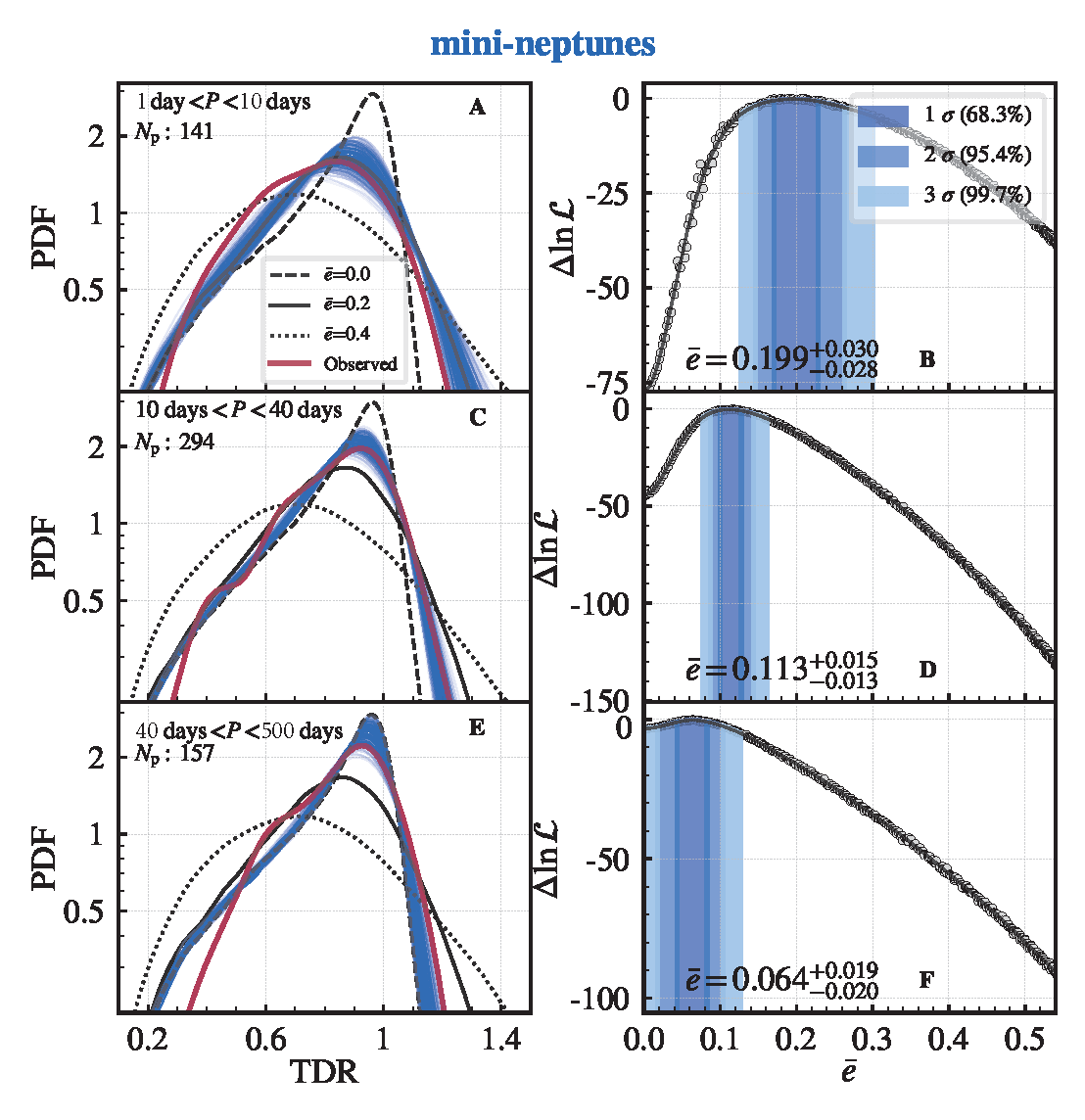}
\captionof{figure}{
\textbf{Same as Fig.~\ref{fs2}, bur for MNs in three orbital period bins.}
}
\label{fs5}
\end{multicols}
\clearpage

\begin{multicols}{2}
\subsection*{\hypertarget{s2}{\emph{P--e} relation from individual measurement}}
\subsubsection*{\hypertarget{s2.1}{The comparison NEA sample}}
\noindent We collected data of transiting small exoplanets (SEs and MNs) from NEA\cite{2013PASP..125..989A,NEA2024} on 6th June 2024 using five criteria shown in Table~\ref{ts2}.
We explain these criteria below.

Criterion 1 selects small exoplanets similar to those in fiducial Kepler sample.
The orbital eccentricities of exoplanets in the NEA catalog are mostly determined using the radial velocity (RV) or transit timing variation (TTV) methods.
If an exoplanet lacks a reliable mass measurement (e.g. has only mass upper limits), its eccentricity could also be unreliable.
Therefore, we require that all exoplanets in our NEA sample have dynamical mass measurements.
We also apply a planet mass upper limit of 13 Jupiter masses ($m<13~M_{\rm J}$) to exclude brown dwarfs with high surface gravities, such as Kepler-2002 b.

Criterion 2 applies an orbital eccentricity cut to select exoplanets with eccentricity measurements. 
The exoplanets that have both zero orbital eccentricities and no upper eccentricity error bars ($e=0$ and without $e_{\rm upper}$) are excluded, because eccentricities are usually fixed to zero when there are no reliable constraints on the eccentricity measurements. 

Criterion 3 is a stellar parameter cut, which is identical to that of the fiducial Kepler sample.

Criterion 4 and criterion 5 are uncertainty cuts on the orbital eccentricity, which account for both the precision and size of the NEA sample.
We examine the effect of different uncertainty cuts below.

After applying these cuts, the comparison NEA sample contains 69 exoplanets in 42 systems (Fig.~\ref{f2}A).
This comparison samples are provided as data S2\cite{zenodo}
\end{multicols}

\begin{table}[!ht]
\centering
\caption{
\textbf{Selection cuts for the comparison NEA sample.}
The first column is the sample selection criterion we add at each step to select exoplanets from the full dataset, and the second column shows the number of exoplanets remaining in the sample after applying the corresponding criterion.
$e$, $e_{\rm upper}$, $e_{\rm lower}$ are the orbital eccentricity, the upper and lower eccentricity uncertainties of the exoplanet, respectively. 
In step 1, we only include exoplanets detected by transit methods to ensure the reliability of their radii measurements.
The exoplanets with masses or radii derived from $m$--$R$ relation, without mass uncertainties or with only mass upper limits are also excluded in step 1.
In step 3, we also remove circumbinary exoplanet systems, such as Kepler-1661(AB) and Kepler-47(AB).
}
\begin{tabular}{lc}
\\
\hline
\multicolumn{1}{c}{Criterion}        & Number of exoplanets \\
\hline
0. NEA database on 6th June 2024          & 5638         \\
1. $0<R<4~R_\oplus$ and $0<m<13~M_{\rm J}$  & 415 \\
2. $0 < e_{\rm upper} <1$ and $0 \leq e_{\rm lower} <1$ and $e-e_{\rm lower}\geq 0$ and $e+e_{\rm upper}<1$ & 150  \\
3. $4700~{\rm K} < T_{\rm eff} < 6500~{\rm K}$ and $\log_{10}(g\cdot{\rm s}^2\cdot{\rm cm}^{-1})>4$ & 90  \\
4. Absolute eccentricity error cut ($\varepsilon<0.12$): $(e_{\rm upper} +  e_{\rm lower})/2 <0.12$ & 76 \\
5. Relative eccentricity error cut ($\eta<120\%$): ${(e_{\rm upper} + e_{\rm lower})/2} <1.2e$ & 69   \\
\hline
\end{tabular}
\label{ts2}
\end{table}

\begin{multicols}{2}
\subsubsection*{\hypertarget{s2.2}{Effect of eccentricity uncertainty cut}}
\begin{figure}[!ht]
\centering
\includegraphics[width=0.85\textwidth]{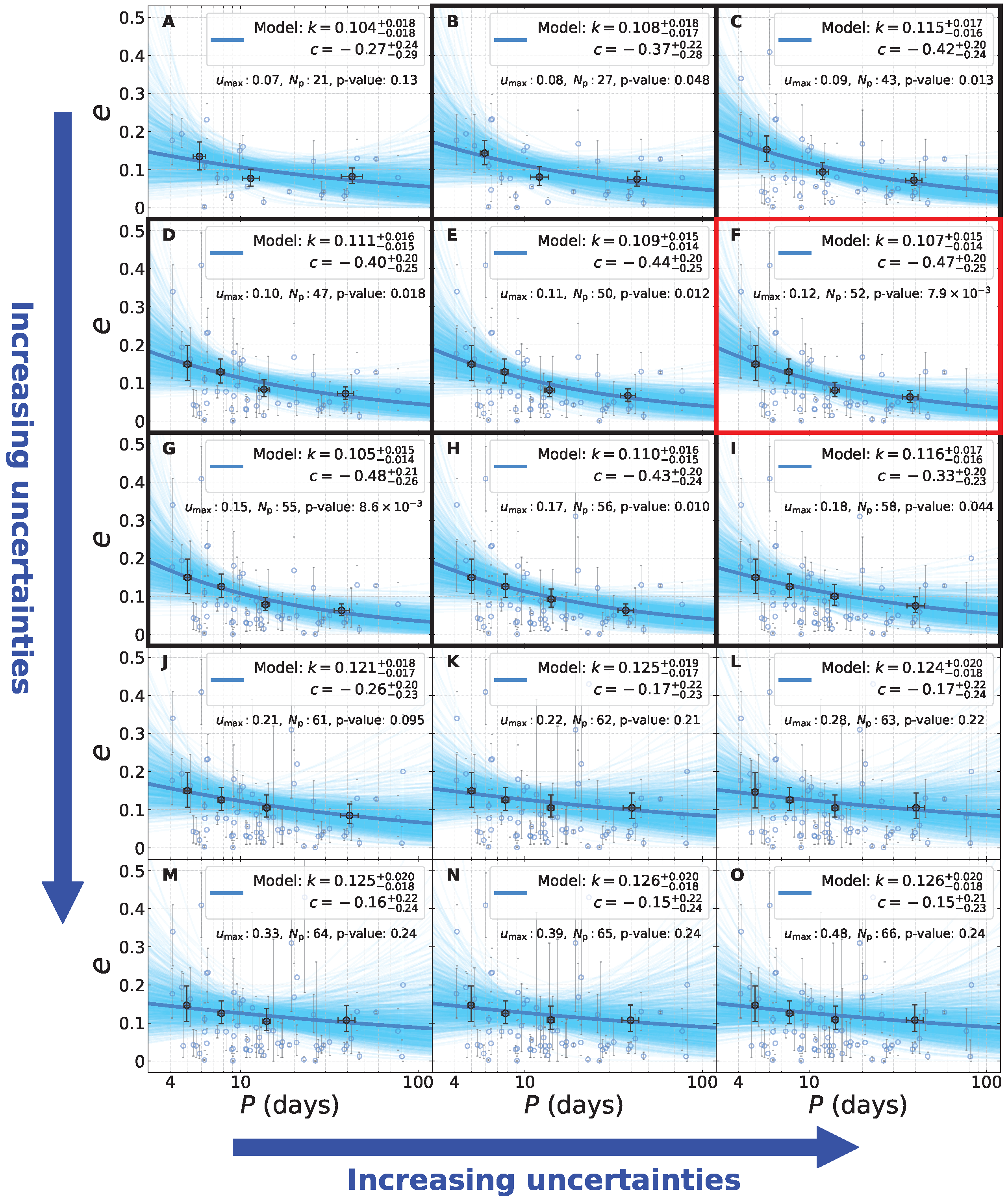}
\caption{
\textbf{Same as Fig.~\ref{f3}B, but for NEA samples with varying eccentricity uncertainty.}
From panel (A) to O, eccentricity uncertainty cuts ($u_{\rm max}$) varying from 0.07 to 0.48.
Each panel is labeled with the corresponding uncertainty upper limit ($u_{\rm max}$), the number of exoplanets (with $P>4$~days) in the samples ($N_{\rm p}$) and the p-value (the fraction of $c\geq0$).
The red outline (panel F) indicates the result of our comparison NEA sample (with $u_{\rm max}=0.12$), while the black outlines (panel B-E \& G-I) indicate the samples with p-value $<0.05$.
}
\label{fs8}
\end{figure}

\noindent For criteria 4 and 5 in the comparison NEA sample, we adopt $\varepsilon < 0.12$ and $\eta < 120\%$, where $\varepsilon$ is the absolute eccentricity error and $\eta$ is the relative eccentricity error.
To investigate the effect of these eccentricity uncertainty cuts, we investigate 14 alternative samples by modifying criterion 4 and criterion 5 to $\varepsilon < \varepsilon_{\rm max}$ and $\eta < \eta_{\rm max}$, with $\varepsilon_{\rm max}$ (the upper limit of $\varepsilon$) varying from 0.07 to 0.48 and $\eta_{\rm max}$ (the upper limit of $\eta$) from 70\% to 480\%.
We define a dimensionless number $u$ to quantify the combined uncertainty, where:
\begin{equation}
\label{es9}
u = 0.5\varepsilon + 0.05\eta.
\end{equation}
Among our 15 samples (including the comparison NEA sample with $u_{\rm max}=0.12$), $u_{\rm max}$ (the upper limit of $u$) varies from 0.07 to 0.48.  
Because there are insufficient SEs in these NEA samples, we focus only on MNs in the statistical analysis (see main text).
We divide MNs in these samples into multiple bins according to their orbital periods to investigate their $P$--$e$ relations.
We apply orbital period boundaries of (4, 6, 10, 20) days to divide the samples with more than 50 exoplanets (with $u_{\rm max}>0.09$, including our comparison NEA sample) into five bins.
For samples with less than 50 exoplanets (with $u_{\rm max}\leq0.09$), we apply four bins with boundaries of (4, 8, 20) days.
These binning criteria ensure that each data bin contains a similar number of plants, while remaining similar to the binning criteria used in our fiducial Kepler sample.
For the following statistical analysis, similar to our fiducial Kepler sample, we exclude the first bins of these samples to avoid the effect of tidal damping (corresponding to an orbital period cut of $P>4$ days, see Supplementary Text).
Then we bootstrap $10^5$ times to obtain the distributions of mean orbital period and mean orbital eccentricity in each bin.
The median values of $\bar{P}$ and $\bar{e}$ are reported for each bin, with uncertainties corresponding to their 68.3\% confidence intervals, as shown in Fig.~\ref{fs8}.

The bootstrap procedure is constructed as follow:
In each iteration, we first randomly select exoplanets with replacements to form a new sub-sample with the same size as the original bin.
This is equivalent to randomly assigning a weight to each exoplanet in the data bin.
Secondly, we re-sample the eccentricities of each exoplanet according to their reported uncertainties (assuming they follow Gaussian distributions).
To avoid negative eccentricities in data re-sample process, we follow previous work\cite{2017A&A...605A..72L} by using $e{\rm sin}\omega$ and $e{\rm cos}\omega$ (instead of $e$) to build the Gaussian distribution.
After data re-sampling, we calculate the mean orbital period and mean eccentricity of exoplanets in each bin, and fit a power-law model ($e = k\cdot(P/(10~{\rm days)})^c$) to the mean orbital periods and mean eccentricities.
The fractions of samples with $c\geq0$ is taken as the p-value of a $P$--$e$ anti-correlation.
The median and corresponding 68.3\% confidence intervals of the posterior probability distributions of each parameter ($k$ and $c$) from $10^5$ fitting processes are taken as the fitting values and $1~\sigma$ uncertainties.
The results of fitting 15 samples (with $P>4$ days) are shown in Fig.~\ref{fs8}.

Fig.~\ref{fs9} shows the sizes and fitting parameters as a function of the parameterized uncertainties ($u_{\rm max}$) for 15 samples (with $P>4$ days).
The sample sizes and fitting parameters vary with the applied uncertainty cuts.
We find p-value $<0.05$, corresponding to a significance $>2~\sigma$  for a $P$--$e$ anti-correlation appears for $u_{\rm max}\in[0.08,0.18]$.
As the uncertainties grow, the best-fitting power-law index $c$ gradually approaches zero, and the p-value gradually increases beyond 0.05 (significance drops below $2~\sigma$), indicating significance of the $P$--$e$ anti-correlation of MNs becomes weaker and is eventually overwhelmed by the increasing uncertainty (Fig.~\ref{fs8}J-O).
In the other direction, as the uncertainty decreases, the sample size (number of exoplanets) becomes too small to obtain reliable fitting results, causing the p-value to increase as well(Fig.~\ref{fs8}A-C).
However, for sample sizes varying from 27 to 58, the fitting parameter $c$ remains stable (consistent within $1~\sigma$), and the p-value remains $<0.05$ (significance $>2~\sigma$).
We conclude that the $P$--$e$ relation of MNs is detected for eccentricity precision $u < 0.18$ and sample size $N_{\rm p} \gtrsim 30$.
\end{multicols}

\begin{multicols}{2} 
\noindent
\includegraphics[width=0.48\textwidth]{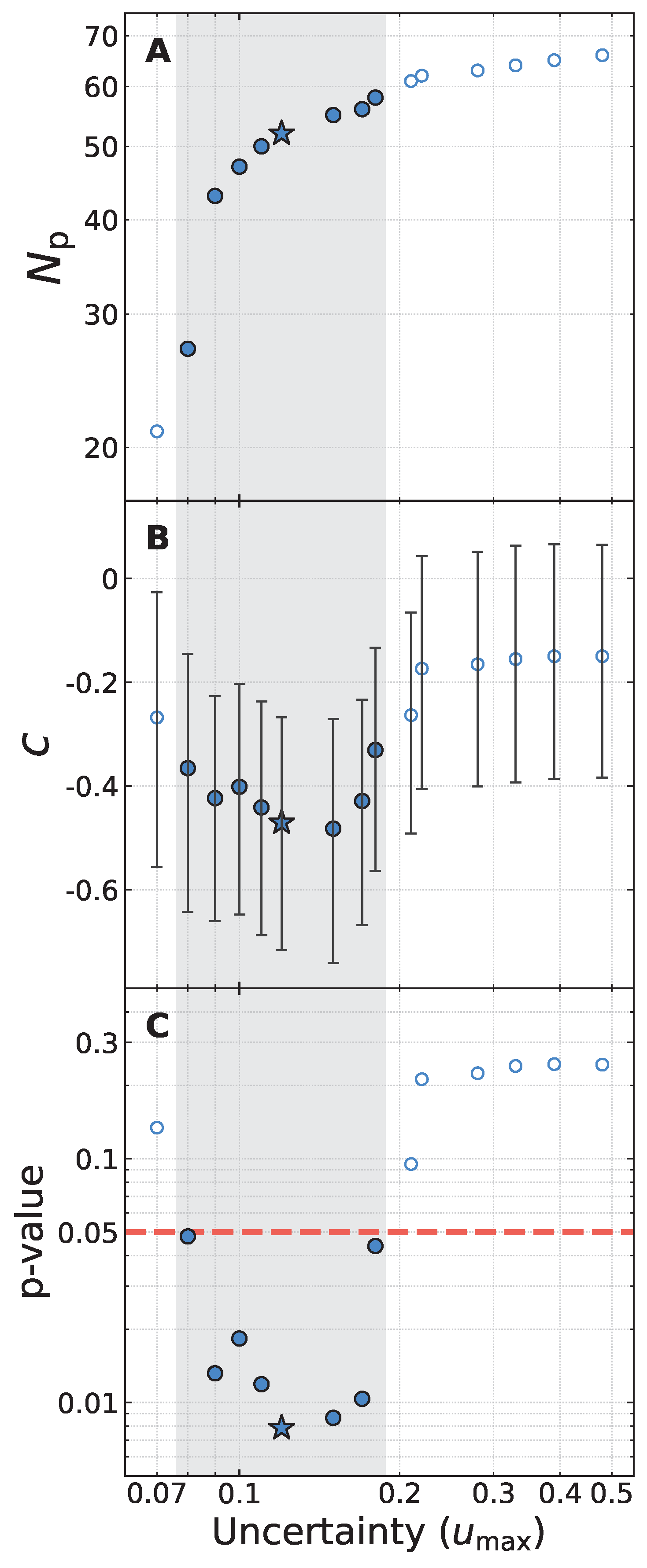}
\captionof{figure}{
\textbf{The NEA sample sizes and fitting results as functions of  eccentricity uncertainty.}
The blue circles and stars show the NEA sample with eccentricity uncertainty cuts ($u_{\rm max}$) varying from 0.07 to 0.48 (corresponding to Fig.~\ref{fs8}A-O).
The stars show our comparison NEA sample (Fig.~\ref{fs8}F), which has the lowest p-value.
The filled circles are the samples with p-values $<0.05$ (Fig.~\ref{fs8}B-E \& G-I), while the open circles are the samples with p-values $>0.05$ (Fig.~\ref{fs8}A \& J-O).
The shaded region indicates the uncertainty range ($0.08 \leq u_{\rm max} \leq 0.18$) for samples with p-values $<0.05$.
{\bf (A)} The sample size as a function of uncertainty.
{\bf (B)} The fitting parameter $c$ (with error bars indicating $1~\sigma$ uncertainties) as a function of uncertainty.
{\bf (C)} The p-value (the fraction of $c\geq0$) as a function of uncertainty.
The red dashed line denotes p-value equal to 0.05 (which corresponds to a significance level of $2~\sigma$).
}
\label{fs9}
\end{multicols}
\clearpage

\begin{multicols}{2}
\section*{Supplementary Text}
\subsection*{\hypertarget{s3}{Theoretical \emph{P--e} relation}}
\subsubsection*{\hypertarget{s3.1}{Planet-planet scattering}}
\noindent The orbital eccentricities of planets can be excited via PPS\cite{1996Sci...274..954R,1996Natur.384..619W}.
The growth of eccentricity through PPS is proportional to the ratio of a planet's surface escape velocity to its orbital velocity\cite{2008ApJ...686..621F,2008ApJ...686..580C}:
\begin{equation}
\label{es10}
e = \beta \frac{v_{\rm esc}}{v_{\rm orb}} = \beta \sqrt{\frac{2ma}{M_\star R}},
\end{equation}
where $\beta$ is a coefficient that quantifies the efficiency of eccentricity excitation, $v_{\rm esc}$ is a planet's escape velocity, and $v_{\rm orb}$ is a planet's orbital velocity.

The theory of PPS can be used to estimate the eccentricity induced by close encounters between planets. 
We adopt an $m$--$R$ relation for SEs\cite{2020A&A...634A..43O}:
\begin{equation}
\label{es11}
\frac{m}{M_\oplus} = 0.90\left(\frac{R}{R_\oplus}\right)^{3.45}, 
\end{equation}
where $M_\oplus$ is the mass of Earth, and apply Kepler's Third Law (assuming $m \ll M_{\star}$), Eq.~\ref{es10} can be rewritten as:
\begin{equation}
\label{es12}
e \approx 0.15\beta \left(\frac{M_{\star}}{M_\odot}\right)^{-\frac{1}{3}} \left(\frac{R}{1.3\;R_\oplus}\right)^{1.2} \left({\frac{P}{10\;{\rm days}}}\right)^{\frac{1}{3}}.
\end{equation}
Eq.~\ref{es12} gives the typical orbital eccentricities of SEs in our fiducial Kepler sample.
From Eq.~\ref{es12} we find that for planets with similar sizes (around host stars with similar $M_{\star}$), the theory of PPS predicts a positive correlation between orbital period and eccentricity, with a power-law index of $c=1/3$.
This value is consistent with our measured value for SEs ($c=0.37^{+0.34}_{-0.20}$).
Fitting Eq.~\ref{es12} to the observed $P$--$e$ distribution of SEs in our fiducial Kepler sample gives $\beta \approx 0.7$ (for $M_{\star}=1~M_\odot$).

However, given that most small exoplanets systems have outer exoplanets more massive than the inner one\cite{2013ApJ...763...41C,2022MNRAS.514.3844C}, we consider the connection between planet mass and orbital period via the MMEN model\cite{2004ApJ...612.1147K,2013MNRAS.431.3444C,2020AJ....159..247D}.
MMEN is an in-situ formation model designed to represent the formation and evolution of SEs.
It assumes that the surface density profile $\Sigma(a)$ of solid planet-building materials in the protoplanetary disk is:
\begin{equation}
\label{es13}
\Sigma(a) = a^\alpha \Sigma_0,
\end{equation}
where $\Sigma_0$ is a scaling coefficient and $\alpha$ is the power-law index of the solid surface density profile.
The exoplanet mass predicted by MMEN is:
\begin{equation}
\label{es14}
m = 2\pi a \Delta a \Sigma(a),
\end{equation}
where $\Delta a$ spans the accretion zone of an exoplanet.
Following previous work\cite{1998Icar..131..171K,2020AJ....159..247D}, we adopt $\Delta a=10R_{\rm Hill}$, where $R_{\rm Hill}$ is the Hill radius of an exoplanet:
\begin{equation}
\label{es15}
R_{\rm Hill} =a \left(\frac{m}{3M_\star}\right)^{\frac{1}{3}}.
\end{equation}
Combining Eq.~\ref{es13}, Eq.~\ref{es14} and Eq.~\ref{es15}, we obtain:
\begin{equation}
\label{es16}
m \approx 288\Sigma_0^{\frac{3}{2}} M_\star^{-\frac{1}{2}} a^{\frac{6+3\alpha}{2}}.
\end{equation}
Combining Eq.~\ref{es10}, Eq.~\ref{es11}, Eq.~\ref{es16}, and apply Kepler's Third Law (assuming $m \ll M_{\star}$), we find:
\begin{equation}
\label{es17}
e \approx 0.09\beta \left(\frac{M_{\star}}{M_\odot}\right)^{-0.47} \left({\frac{P}{10\;{\rm days}}}\right)^{0.42},
\end{equation}
for $\Sigma_0=50~{\rm g \cdot cm^{-2}}\cdot{\rm au}^{-\alpha}$ and $\alpha=-1.75$\cite{2020AJ....159..247D}, where au is the astronomical units.
Eq.~\ref{es17} has an exponent of $c\approx0.42$, which is consistent with our measurement for SEs, and more closely matches the $P$--$e$ relation of SEs after correcting the effect of stellar metallicity (see below).
Fitting Eq.~\ref{es17} to the observed $P$--$e$ distribution of SEs in our fiducial Kepler sample gives $\beta \approx 1$ (for $M_{\star}=1~M_\odot$), indicating efficient eccentricity excitation.
However, in this model, the value of $\beta$ is dependent on $\Sigma_0$ ($\beta \sim \Sigma_0 ^{-0.53}$).
For example, if we adopt a larger $\Sigma_0=100~{\rm g \cdot cm^{-2}}\cdot{\rm au}^{-\alpha}$ derived from RV measurements\cite{2020AJ....159..247D}, we find $\beta \approx 0.7$.

\subsubsection*{\hypertarget{s3.2}{Angular momentum deficit equipartition}}
\noindent In a planetary system that is well spaced, if planets are not in exact resonance, the dynamical evolution is governed by long-range secular interactions among planets.  
In the secular dynamical evolution, the total AMD is conserved while exchanging among different planets\cite{1997A&A...317L..75L,2000PhRvL..84.3240L,2017A&A...605A..72L}, ultimately leading to AMD equipartition among different planets in the planetary systems\cite{1989Icar...77..330W,2011ApJ...735..109W,2017A&A...605A..72L}.
Planets near mean motion resonances (MMR) are expected to reach AMD equipartition more quickly\cite{2011ApJ...735..109W}.

The AMD is defined as the difference in the normal component of angular momentum (with respect to the system invariant plane) that a planet has compared to what it would have if its orbit was circular and coplanar (assuming $m \ll M_{\star}$)\cite{2017A&A...605A..72L}:
\begin{equation}
\label{es18}
{\rm AMD}= m \sqrt{{\rm G}M_{\star}a}(1-\sqrt{1-{e}^2}{\rm{cos}}i),
\end{equation}
where $i$ is the inclination of the planet relative to the system invariant plane.

As the system approaches AMD equipartition, the AMD of each planet in the system becomes nearly equal.
Assuming both orbital eccentricity and inclination are low ($e^2\sim0$ and $i\sim0$), Eq.~\ref{es18} indicates the eccentricity of a planet is:
\begin{equation}
\label{es19}
e \approx \left(\frac{4{\rm AMD_0}^2}{{\rm G}M_{\star}} \right)^{\frac{1}{4}} m^{-\frac{1}{2}}a^{-\frac{1}{4}},
\end{equation}
where $\rm AMD_0$ is the AMD of a planet.
We rewrite Eq.~\ref{es19} as:
\begin{equation}
\label{es20}
e \approx 1.92{\rm AMD_0}^{\frac{1}{2}}({{\rm G}M_{\star}})^{-\frac{1}{3}}m^{-\frac{1}{2}}P^{-\frac{1}{6}},
\end{equation}
by applying Kepler's Third Law (assuming $m \ll M_{\star}$).
From Eq.~\ref{es20}, we find that for planets with similar masses (around host stars with similar $M_{\star}$), AMD equipartition predicts an anti-correlation between orbital period and eccentricity, with a power-law index $c=-1/6$.
This is substantially weaker than our measured value for MNs ($c=-0.45^{+0.13}_{-0.15}$).
Therefore, we further consider the correlation between orbital period and planet mass.

\begin{figure}[!ht]
\centering
\includegraphics[width=1\textwidth]{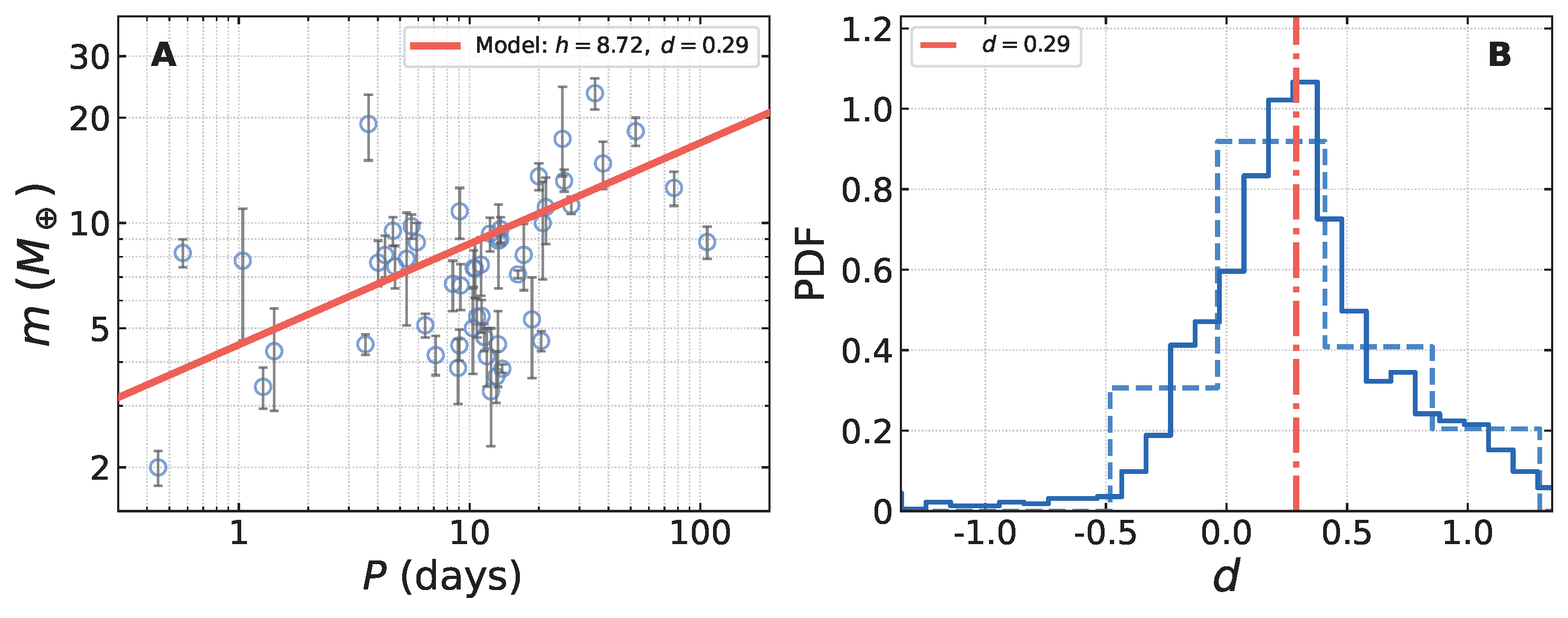}
\caption{
\textbf{The period--mass distribution of MNs from NEA catalog.}
{\bf (A)} The blue circles show planet mass (with error bars indicating $1~\sigma$ uncertainties) as a function of orbital period of exoplanets in multi-exoplanet systems containing at least one MN.
The red line is the median power-law model ($m/M_\oplus=h\cdot(P/(10~{\rm day}))^d$) from the bootstrap re-sampling, which has the parameters listed in the legend.
{\bf (B)} The PDF of power-law index $d$.
The dashed histogram is the original distribution and the solid histogram is the bootstrapping distribution.
The red dot-dashed line indicates the median value ($d=0.29$) of the bootstrapping distribution.
}
\label{fs10}
\end{figure}

Because the formation and evolution of MNs can involve orbital migration, instead of using the MMEN model, we adopt an empirical $P$--$m$ relation from observations.
We collect a sample of multi-planet systems (each containing at least one MN with $R_{\rm valley}<R<4~R_\oplus$) with precise mass measurement in the NEA catalog, with these selection criteria:
\begin{enumerate}
\item Planetary parameter cut: $0<R<4~R_\oplus$, $0<m<13~M_{\rm J}$, and only include exoplanets that are measured by both transit and RV methods.
\item Relative mass uncertainty cut: $(m_{\rm upper} +  m_{\rm lower})<m$, where $m_{\rm upper}$ and $m_{\rm lower}$ are the upper and lower mass uncertainties of the exoplanet, respectively. 
\item Stellar parameter cut: $4700~{\rm K} < T_{\rm eff} < 6500~{\rm K}$, $\log_{10}(g\cdot{\rm s}^2\cdot{\rm cm}^{-1})>4$ , and exclude circumbinary exoplanet systems.
\end{enumerate}
These cuts result in 52 exoplanets in 22 systems in the sample, and their $P$--$m$ distribution is shown in the left panel of Fig.~\ref{fs10}A.

We fitted the $P$--$m$ distribution of exoplanets in each multi-planet system with a power-law model: $m/M_\oplus=h\cdot(P/(10~{\rm day}))^d$, where $h$ (amplitude) and $d$ (index) are the fitting parameters.
The distribution of 22 power-law indices $d$ is shown in Fig.~\ref{fs10}B.
To account for the uncertainties of the mass measurements, we obtain a alternative distribution of $d$ by performing bootstrap  re-sampling.
Specifically, for each system, we re-sample the planet mass 100 times according to their reported error uncertainties (by assuming they follow Gaussian distributions).
In each re-sample process, we fitted a power-law model to the $P$--$m$ distribution of exoplanets in each multi-planet system to obtain 100 fitting parameters ($h$ and $d$) for each system.
The distribution of the resulting 2200 values of $d$ is also shown in Fig.~\ref{fs10}B.

Most of these multi-planet systems follow a positive $P$--$m$ correlation ($d>0$), and the median $P$--$m$ correlation (from the bootstrapping distribution) is:
\begin{equation}
\label{es21}
\frac{m}{M_\oplus} = 8.72\left(\frac{P}{10\;{\rm days}}\right)^{0.29}.
\end{equation}
The median power-law index ($d=0.29$) of the observed $P$--$m$ correlation of MNs is similar to that predicted by MMEN model ($d=0.25$, calculated using Eq.~\ref{es16} with $\alpha=-1.75$)\cite{2020AJ....159..247D}.
Combining Eq.~\ref{es21} with Eq.~\ref{es20}, we obtain:
\end{multicols}

\begin{equation}
\label{es22}
e \approx 0.20 \left(\frac{\rm AMD_0}{\rm AMD_{Neptune}}\right)^{\frac{1}{2}} \left(\frac{M_{\star}}{M_\odot}\right)^{-\frac{1}{3}}  \left({\frac{P}{10\;{\rm days}}}\right)^{-0.31},
\end{equation}

\begin{multicols}{2}
\noindent where $\rm AMD_{Neptune}$ is the AMD of Neptune in the Solar system.
The exponent in Eq.~\ref{es22} is negative, which implies an anti-correlation between orbital period and eccentricity.
This predicted power-law index of $c\approx-0.31$ is $1.1~\sigma$ larger than our measured value for MNs ($c=-0.45^{+0.13}_{-0.15}$).
This slight difference could be explained by the effect of stellar metallicity (see below).
Fitting Eq.~\ref{es22} to the observed $P$--$e$ distribution of MNs in our fiducial Kepler sample gives ${\rm AMD_0}\approx 0.65\rm AMD_{Neptune}$ (for $M_{\star}=1~M_\odot$).
\end{multicols}

\begin{multicols}{2}
Eq.~\ref{es22} was derived assuming that all MN systems have similar outer planetary systems (same $P$--$m$ relations, Eq.~\ref{es21}).
However, real MN systems likely have different outer systems and therefore would exhibit varying amount of available AMD (there may also be systematic correlations between the properties of MNs and the available AMD from the outer system), causing their $P$--$e$ relations to deviate from Eq.~\ref{es22}.

\subsubsection*{\hypertarget{s3.3}{Effect of tidal damping}}
\noindent In principle, planets with shorter orbital periods should undergo stronger gravitational tidal interactions, which act to circularize their orbits.
If we consider only the effect of the gravitational tide raised on the planet by the star, the equilibrium tide can be approximately described by a constant $Q$ model (where $Q$ is tidal quality factor).
By assuming $e^2\sim0$ and ignoring the stellar tides, we have\cite{1966Icar....5..375G,2008ApJ...678.1396J}:
\begin{equation}
\label{es23}
{\frac{1}{e}} \frac{{\rm d}e}{{\rm d}t}  = -\frac{1}{\tau_e},
\end{equation}
where $\tau_e$ is the tidal circularization timescale:
\begin{equation}
\label{es24}
\tau_e = Q'{\frac{P}{21\pi}}\frac{m}{M_{\star}}\left({\frac{a}{R}}\right)^5,
\end{equation}
$Q'\equiv Q/k_2$ is the modified tidal quality factor and $k_2$ is the second Love number.
The smaller the $Q'$, the stronger the tidal damping effect.
The values of $Q'$ for small exoplanets (SEs and MNs) are highly uncertain  \cite{2017AJ....153...86M,2018AJ....155..157P,2023ApJ...958L..21L}.
It could range from $\sim 10^2$ (rocky planet)\cite{2016CeMDA.126..145L} to $\sim 10^5$ (Neptunian planet)\cite{2008Icar..193..267Z}.
It has been predicted that SEs have a smaller $Q'$ ($\sim 10^3$, closer to the rocky value) than MNs ($\sim 10^4$, closer to the Neptunian value)\cite{2023ApJ...958L..21L}.

To evaluate the tidal circularization timescales of SEs and MNs, we adopt $m$--$R$ relations\cite{2020A&A...634A..43O}:
\begin{equation}
\label{es25}
\frac{m}{M_\oplus}  = 
\begin{cases}
1.74\left(\frac{R}{R_\oplus}\right)^{1.58}~{\rm for~MNs}  ~(R>R_{\rm  valley}) \\
0.90\left(\frac{R}{R_\oplus}\right)^{3.45}~{\rm for~SEs} ~(R<R_{\rm  valley})
\end{cases}
.
\end{equation}
Combining Eq.~\ref{es24} with Eq.~\ref{es25}, we obtain:
\end{multicols}

\begin{equation}
\label{es26}
\tau_e = 
\begin{cases}
4.5~{\rm Gyrs} \; \frac{Q'}{10^4}  \left(\frac{M_\star}{M_\odot}\right)^{\frac{2}{3}}  \left({\frac{P}{6\;{\rm days}}}\right)^{\frac{13}{3}} \left({\frac{R}{2.5~R_\oplus}}\right)^{-3.42}, ~{\rm for~MNs}  ~(R>R_{\rm  valley}) \\
3.6~{\rm Gyrs} \; \frac{Q'}{10^3}  \left(\frac{M_\star}{M_\odot}\right)^{\frac{2}{3}}  \left({\frac{P}{6\;{\rm days}}}\right)^{\frac{13}{3}} \left({\frac{R}{1.3~R_\oplus}}\right)^{-1.55}, ~{\rm for~SEs}  ~(R<R_{\rm  valley})
\end{cases}
.
\end{equation}

\begin{multicols}{2}
\noindent Eq.~\ref{es26} gives the typical tidal circularization timescales of SEs and MNs in the fiducial Kepler sample. 
SEs (MNs) exhibit stronger (weaker) self-gravitation than rocky (Neptunian) planets, which will lead to weaker (stronger) tidal damping.
Taking into account this size effect\cite{2012ApJ...746..150E}, we consider a larger $Q'$ of SEs compared to rocky planets and a smaller $Q'$ of MNs compared to Neptunian planets in Eq.~\ref{es26}, following previous work\cite{2023ApJ...958L..21L}.

The median stellar ages of SEs and MNs in our fiducial Kepler sample are 4.1 and 4.7 Gyrs, respectively, so we expect exoplanets with short orbital periods ($P\lesssim6$ days) to undergo prominent tidal circularization.
From Fig.~\ref{f1}B-C, we find that tidal damping is prominent for exoplanets with short orbital periods ($\lesssim$4−6 days), affecting a larger fraction (4 of 9 bins) of SEs than MNs (1 of 9 bins) in our fiducial Kepler sample.
We infer that the eccentricity distributions of both SEs and MNs in these short-period systems are consistent with the effects of tidal damping.

For MNs, the lower eccentricity of the first bin in Fig.~\ref{f1}B (with $P<4$ days) appears to violate the $P$--$e$ anti-correlation.
We interpret this as due to effect of tidal damping.
The eccentricity of the second bin in Fig.~\ref{f1}B (with $4~{\rm days}<P<6~{\rm days}$) appears unaffected by tidal damping, and remains consistent with the overall $P$--$e$ anti-correlation (see below).
We attribute this to the AMD transfer.
In the AMD equipartition scenario, AMD can be continuously transferred to the inner exoplanets over a timescale $\tau_{\rm AMD}\sim100$ Myrs\cite{2011ApJ...735..109W}, maintaining their eccentricities.
According to Eq.~\ref{es26}, the tidal circularization timescales of MNs with $P\in$ (4,6) days are $\sim$ Gyrs, much longer than their AMD equipartition timescales ($\tau_e \gg \tau_{\rm AMD}$).
This could explain the weaker tidal damping in the second bin, which is consistent with the overall $P$--$e$ anti-correlation.
Therefore, we conclude that an initial eccentricity distribution anti-correlated with orbital periods is consistent with our observation, while the lower eccentricities in short periods can be explained by the effect of tidal damping.

For SEs, the first four bins (with $P<6$ days) all display low eccentricities: $\bar{e}\lesssim0.1$ (Fig.~\ref{f1}C).
This indicates that tidal damping plays a role in the positive $P$--$e$ correlation of SEs.
If SEs initially exhibited eccentricities independent of orbital periods ($e=e_0$), tidal damping would produce a nearly bimodal eccentricity distribution ($e\sim0$ in short orbital period regime, and $e\sim e_0$ in long orbital period regime).
Instead, we observe a gradually increasing trend in the $P$--$e$ distribution of SEs (Fig.~\ref{f3}C).
Therefore, we find that an initial eccentricity distribution positively correlated with orbital periods, in line with the predictions of the PPS model, is more consistent with the observed characteristics of SEs. 
However, the $\sim~2.4\sigma$ statistical significance of this correlation is low, particularly when short-period ($P\lesssim6$ days) exoplanets are excluded from the analyses (see below).

Furthermore, the tentatively positive $P$--$e$ correlation does not exclude secular interactions in SE systems, rather it indicates that they have not reached AMD equipartition. 
Secular interactions tend to transfer AMD to inner exoplanets within the systems, thereby increasing their orbital eccentricities.
However, stronger tidal effects (due to shorter orbital periods and smaller tidal quality factors compared to MNs) simultaneously damp the eccentricities of these inner SEs, causing them to migrate inward (a process that further strengthens the tidal effect).
This would cause SEs with short orbital periods in mature systems to exhibit low eccentricities.
A potential evidence of this ongoing excitation-damping process is that observed SEs with short orbital periods show low but non-zero eccentricities\cite{2020A&A...635A..37C,2024AJ....168..115B}.
Therefore, we propose that the positive $P$--$e$ relation of SEs is consistent with the combined effect of PPS, tidal damping and secular interactions. 
While the moderate eccentricities in long orbital periods (where the effect of tidal damping and secular interactions can be ignored) can be explained by PPS, the lower eccentricities in short periods (where the effect of PPS can be ignored) agree with the joint effect of tidal damping and secular interactions

\subsection*{\hypertarget{s4}{Data point dropout test}}
\begin{figure}[!ht]
\centering
\includegraphics[height=0.84\textheight]{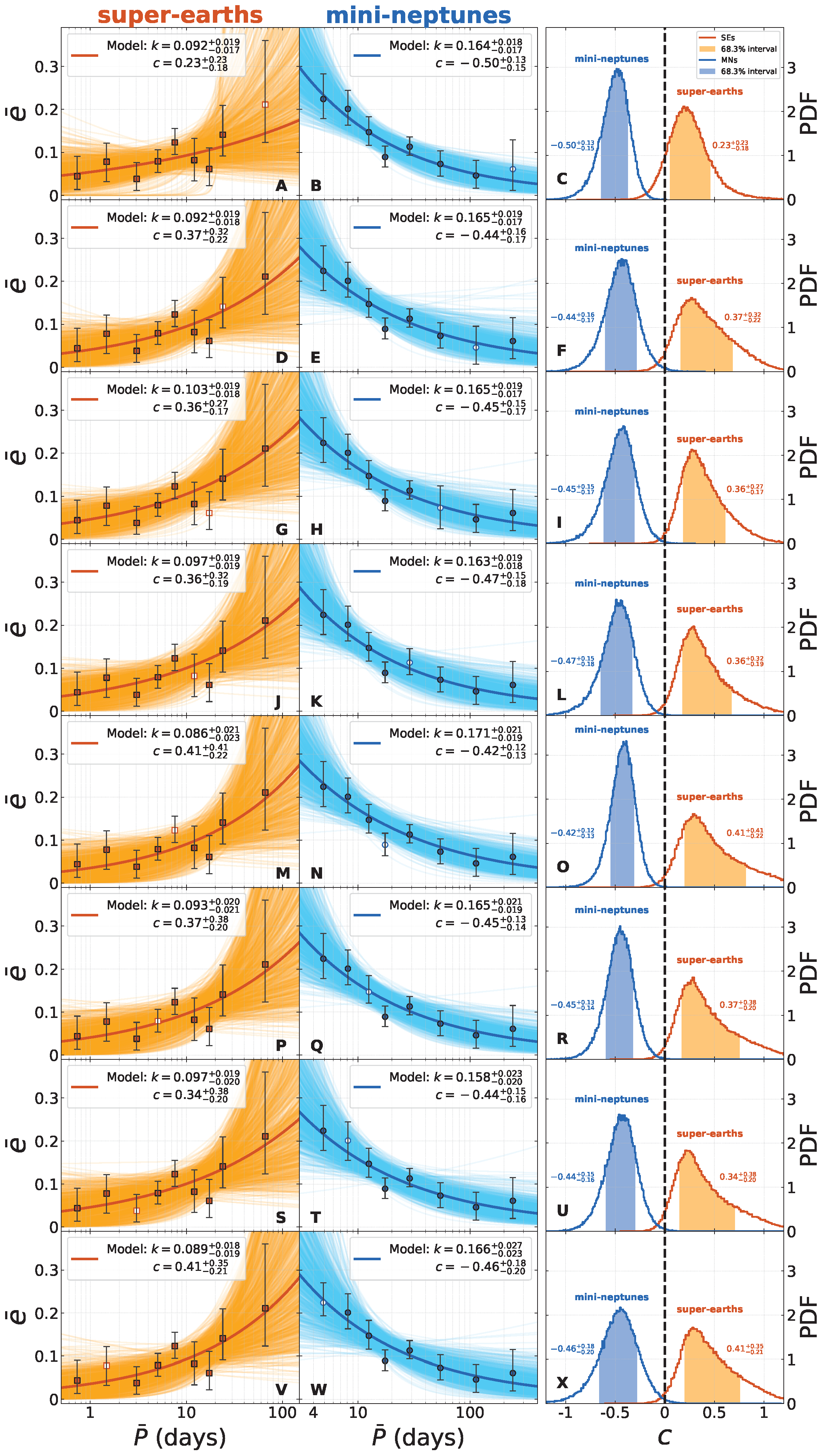}
\caption{
\textbf{Data point dropout test of the Kepler samples.}
Same as Fig.~\ref{f3}A, C\&D, but each row shows a sample with one bin removed (open square or open circle).
}
\label{fs11}
\end{figure}

\begin{figure}
\centering
\includegraphics[width=1\textwidth]{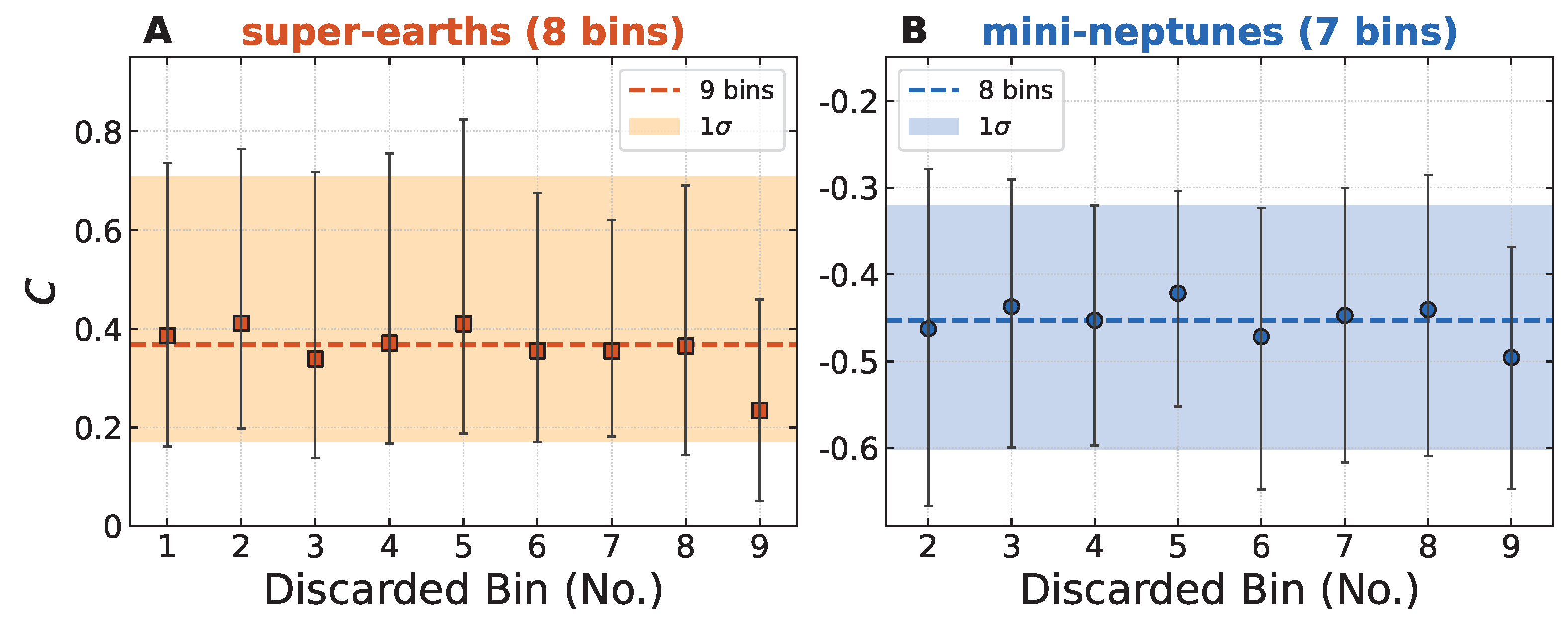}
\caption{ 
\textbf{The power-law index as a function of the discarded bin number.}
The orbital period increases from bin No.1 to bin No.9.
The dashed lines and shaded regions indicate the values of $c$ and the $1~\sigma$ uncertainty from the result shown in Fig.~\ref{f3}A\&C (9 bins for SEs and 8 bins for MNs).
{\bf (A)} The squares are the power-law indices of $P$--$e$ relations (with error bars indicating $1~\sigma$ uncertainties) for SEs.
{\bf (B)} Same as panel (A), but for MNs.
}
\label{fs12}
\end{figure}

\noindent To assess the impact of individual data points on the $P$--$e$ relations of SEs and MNs, we conduct a data point dropout test.
Specifically, we perform the same power-law fitting procedures for SEs and MNs in our fiducial Kepler sample but exclude one data bin each time.
The fitting results are shown in Fig.~\ref{fs11}, and we plot the power-law index $c$ as a function of the discarded bin number in Fig.~\ref{fs12}.

Fig.~\ref{fs11} has eight examples, each dropping one data point.
We find the results for MNs are not sensitive to the discarded data: the $P$--$e$ anti-correlation significance varies from $2.7~\sigma$ to $4.3~\sigma$ (the fractions of $c\geq0$ range from $7.5\times13^{-3}$ to $2.0\times13^{-5}$).
The power-law indices $c$ of MNs remain consistent, regardless of which bin we exclude (Fig.~\ref{fs12}B).
For SEs, in most cases the significance of $P$--$e$ positive correlation is about $2~\sigma$ (the fractions of $c\leq0$ range from 0.035 to 0.013), except for the case of dropping the outermost bin (Fig.~\ref{fs11}A), for which the confidence level reduces to $1.6~\sigma$ (the fraction of $c\leq0$ is 0.10).
The power-law indices $c$ of SEs remain consistent within $1~\sigma$, regardless of which bin we exclude (Fig.~\ref{fs12}A).
\end{multicols}

\begin{multicols}{2}
\subsection*{\hypertarget{s5}{Effect of changing radius boundary}}
\noindent In our Kepler and NEA samples, we use the empirical location of the radius valley (Eq.~\ref{es8}) as the boundaries between SEs and MNs, which is a function of both orbital period and stellar mass.
Here, we test how different choices of this boundary will influence the measured $P$--$e$ relations for both populations.
We construct 20 additional samples based on our fiducial Kepler sample by varying the boundaries between SEs and MNs ($R_{\rm boundary}$), as well as the upper radius limits of MNs ($R_{\rm max}$).
Specifically, we vary $R_{\rm boundary}$ from 1.7 to 2.1~$R_\oplus$ ($1.9\pm0.2~R_\oplus$, without any dependence on orbital periods or stellar mass), and $R_{\rm max}$ from 4 to 6~$R_\oplus$ (close to the Neptune desert\cite{2011ApJ...727L..44S,2016A&A...589A..75M,2024A&A...689A.250C}).
For all these samples, we perform the same data binning and fitting procedures as our fiducial Kepler sample.

\begin{figure}[!ht]
\centering
\includegraphics[width=1\textwidth]{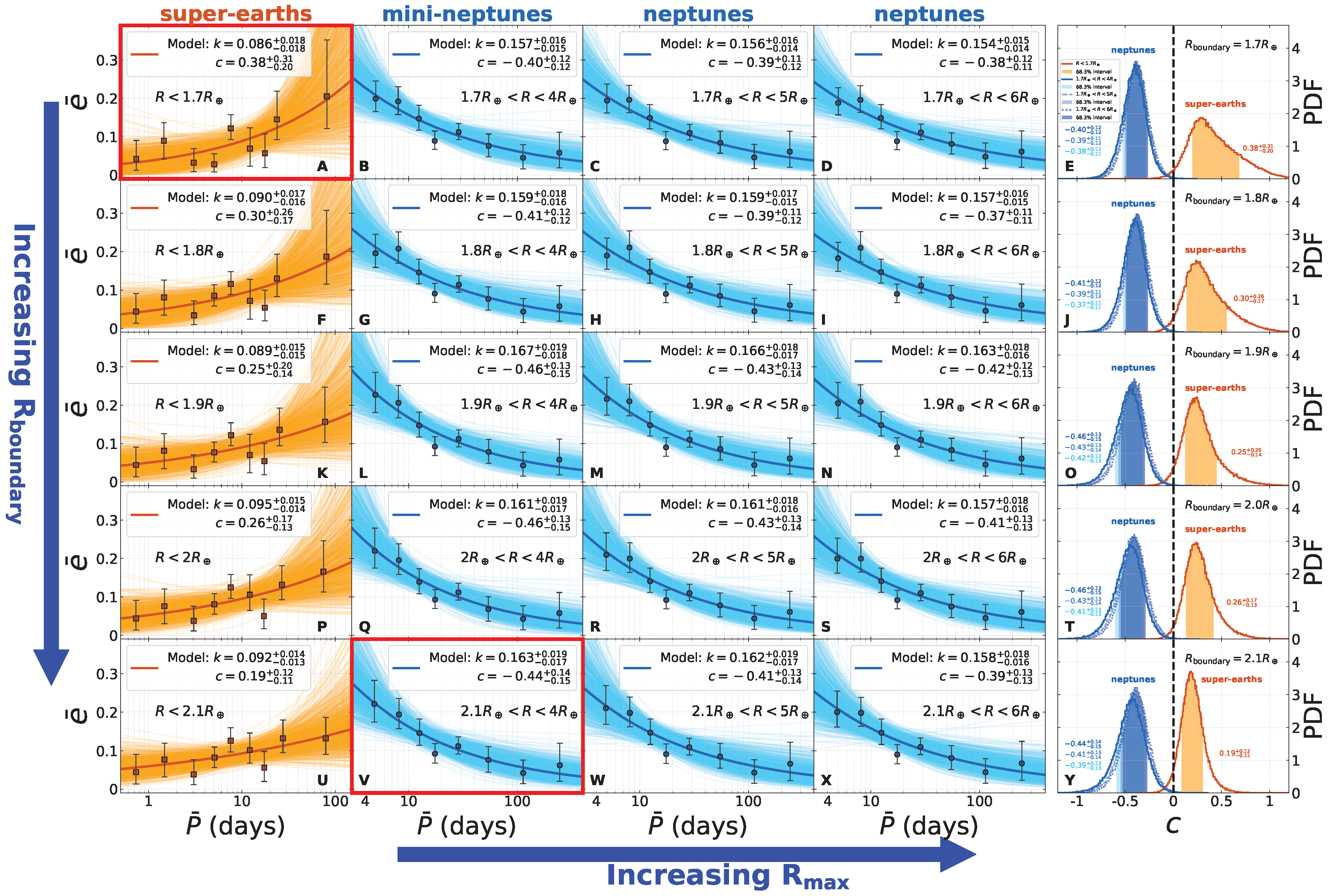}
\caption{
\textbf{Same as Fig.~\ref{f3}A, C\&D, but for Kepler samples with different radius boundaries.}
The radius boundary ($R_{\rm boundary}$) increases from top to bottom, and the upper radius limit of MNs ($R_{\rm max}$) increases from left to right.
Red outlines highlight two samples that exclude exoplanets inside the radius valley ($1.9\pm0.2~R_\oplus$)
}
\label{fs13}
\end{figure}

Fig.~\ref{fs13} shows that the different $P$--$e$ relations of the two populations persist despite the changes in radius boundaries and upper limits.
The power-law indices $c$ of the $P$--$e$ relations remain distinct between SEs and MNs (the fifth column of Fig.~\ref{fs13}).

\begin{figure}[!ht]
\centering
\includegraphics[width=1\textwidth]{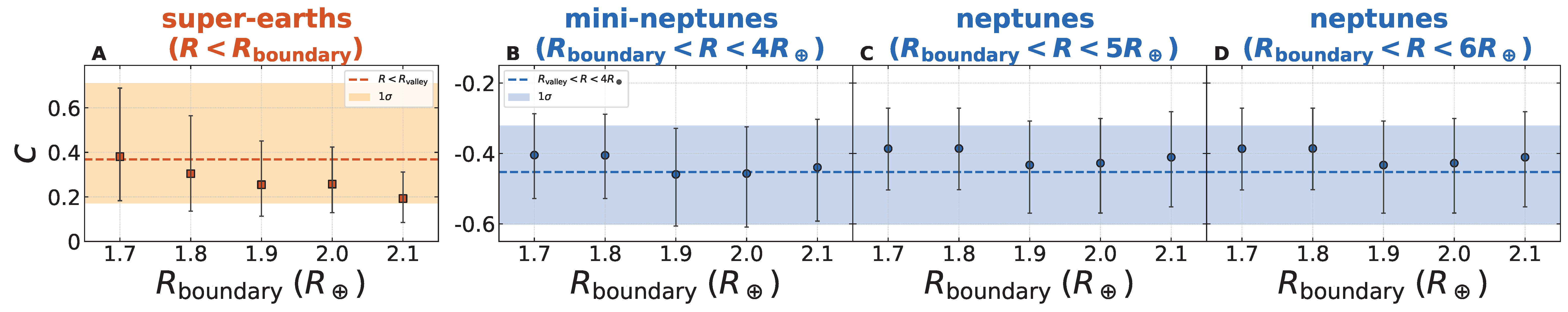}
\caption{
\textbf{The power-law index as a function of the radius boundary.}
The dashed lines and shaded regions indicate the values of $c$ and the $1~\sigma$ uncertainty from the result shown in Fig.~\ref{f3}A\&C ($R<R_{\rm valley}$ for SEs and $R_{\rm valley}<R<4R_\oplus$ for MNs).
{\bf (A)} The squares are the power-law indices of $P$--$e$ relations (with error bars indicating $1~\sigma$ uncertainties) for SEs ($R<R_{\rm boundary}$).
{\bf (B)} Same as panel (A), but for MNs($R_{\rm boundary}<R<4R_\oplus$).
{\bf (C)} Same as panel (A), but for neptunes ($R_{\rm boundary}<R<5R_\oplus$).
{\bf (D)} Same as panel (A), but for neptunes ($R_{\rm boundary}<R<6R_\oplus$).
}
\label{fs14}
\end{figure}

By investigating the power-law index $c$ as a function of the radius boundary $R_{\rm boundary}$, we find that the positive $P$--$e$ relation of SEs gradually weakens as the boundary increases (Fig.~\ref{fs14}A).
We attribute this to the contamination of MNs in the SE sample: as the boundary expands from 1.7 to 2.1~$R_\oplus$, an increasing number of MNs are included in the SE sample.
Consequently, because MNs exhibit an anti-correlation between orbital periods and eccentricities, the positive $P$--$e$ correlation of SEs is weakened.

For MNs, the power-law indices change by $<1~\sigma$ regardless of changes in radius boundary and upper limit (Fig.~\ref{fs14}B-D).
The slightly increase in $c$ (weakening of the $P$--$e$ anti-correlation), as the boundary decreases to less than $1.9~R_\oplus$, is likely caused by contamination of the sample by SEs.

Fig.~\ref{fs13}A shows a SE sample with $R<1.7~R_\oplus$ and Fig.~\ref{fs13}V shows a MN sample with $2.1~R_\oplus<R<4~R_\oplus$.
By removing exoplanets in the radius valley ($1.9\pm0.2~R_\oplus$), these two samples are well separated in the planetary radius distribution, with the least mutual contamination.
We find the $P$--$e$ relations of these two samples are consistent with those of the fiducial Kepler samples shown in Fig.~\ref{f3}.
We conclude that the distinct $P$--$e$ relations between SEs and MNs are not sensitive to the adopted radius boundaries.
\end{multicols}

\begin{multicols}{2}
\subsection*{\hypertarget{s6}{Effect of orbital period cut}}
\noindent We apply an orbital period cut of $P>4$ days to MNs in our sample to avoid tidal damping effects.  
In this section, we test the effect of this orbital period cut on the observed orbital $P$--$e$ relations for SEs and MNs.

As discussed above, the typical tidal circularization timescales of MNs with $P<4$ days are $<1$ Gyr.
Therefore, we expect these exoplanets to exhibit lower orbital eccentricity.
Although MNs with $P\in$ (4,6) days also have tidal circularization timescales comparable to the system ages, excluding these exoplanets dose not affect the observed $P$--$e$ anti-correlation of MNs.
Fig.~\ref{fs11}W shows that fitting the last 7 bins (corresponding to an orbital period cut of $P>6$ days) of MNs in our fiducial Kepler sample yields a power-law index $c$ of $-0.46^{+0.18}_{-0.20}$, which is consistent with the results in Fig.~\ref{f3}.
\end{multicols}

\begin{figure}[!ht]
\centering
\includegraphics[width=1\textwidth]{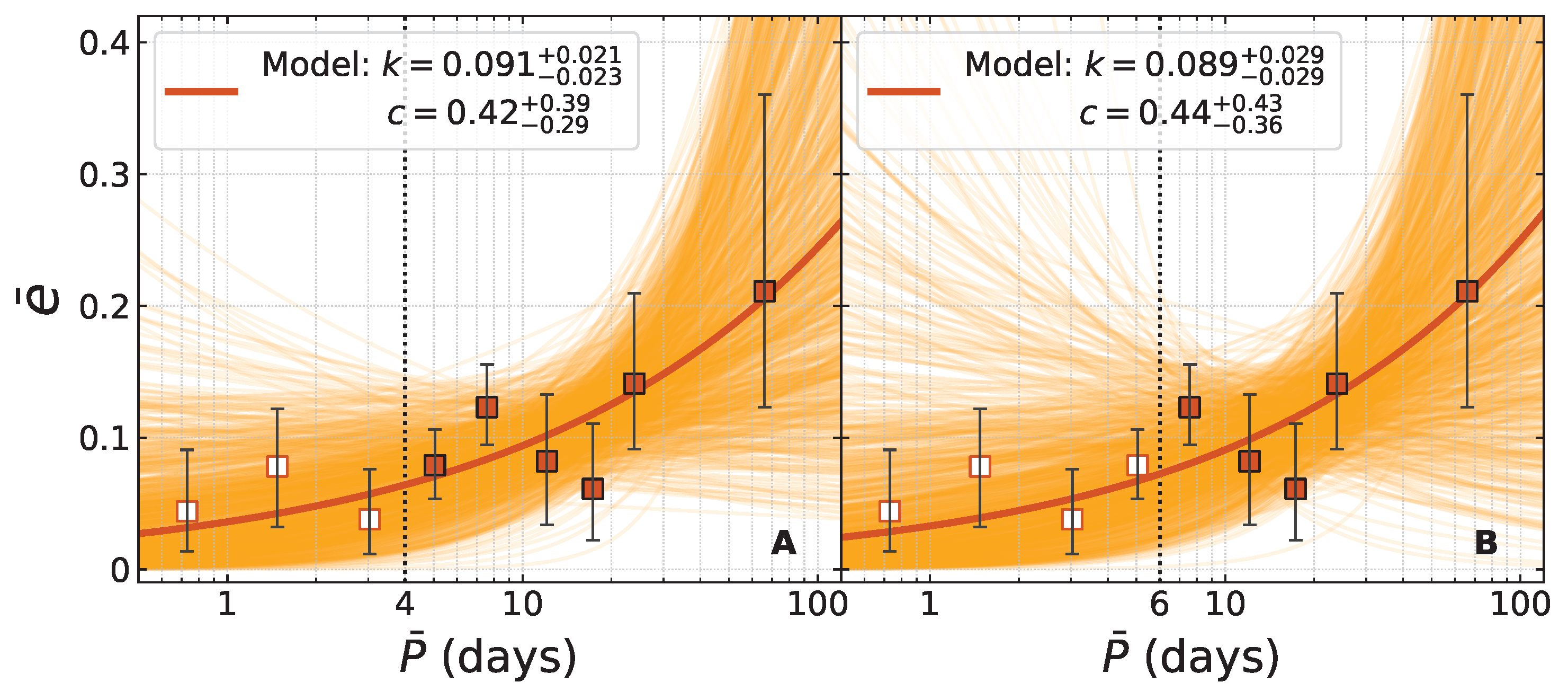}
\caption{
\textbf{Same as Fig.~\ref{f3}C, but for SEs with different orbital period cuts.}
The open squares are the excluded data bins.
The dotted lines indicate the orbital period thresholds.
{\bf (A)} SEs in Kepler sample with an orbital period cut of $P>4$ days.
{\bf (B)} SEs in Kepler sample with an orbital period cut of $P>6$ days.
}
\label{fs15}
\end{figure}

\begin{multicols}{2}
For SEs, as discussed above, the low orbital eccentricities in short periods are consistent with our expectations for tidal damping.
Here we investigate whether applying an orbital period cut affects the observed $P$--$e$ relation of SEs.
We re-perform the fitting procedures for SEs in our fiducial Kepler sample, utilizing only the last six (five) bins, which corresponds to an orbital period cut of $P>4$ days ($P>6$ days).
Fig.~\ref{fs15} shows the best-fitting power-law indices are $c=0.42^{+0.39}_{-0.29}$ for the 4-day period cutoff and $c=0.44^{+0.43}_{-0.36}$ for the 6-day cutoff.
These values are consistent with $c=0.37^{+0.34}_{-0.20}$ in Fig.~\ref{f3}.
However, the exclusion of a substantial data fraction (3 of 9 bins for the 4-day cut and 4 of 9 bins for the 6-day cut) increases the uncertainty on $c$.
Consequently, the fraction of $c>0$ from the posterior probability distributions decreases from $98.2\%$ to $93.0\%$ and $89.0\%$, while the probability of posterior overlap between SE and MN populations increases from $1.8\times13^{-4}$ to $2.3\times13^{-3}$ and $1.2\times13^{-2}$ respectively.
Correspondingly, the statistical significance of the positive $P$--$e$ correlation weakens from $\sim2.4~\sigma$ to $\sim1.8~\sigma$ and $\sim1.6~\sigma$, and the statistical significance of the distinct $P$--$e$ relations of SEs and MNs decreases from $\sim3.7~\sigma$ to $\sim3.0~\sigma$ and $\sim2.5~\sigma$.

\subsection*{\hypertarget{s7}{Effect of SNR Cut}}
\begin{figure}[!ht]
\centering
\includegraphics[width=1\textwidth]{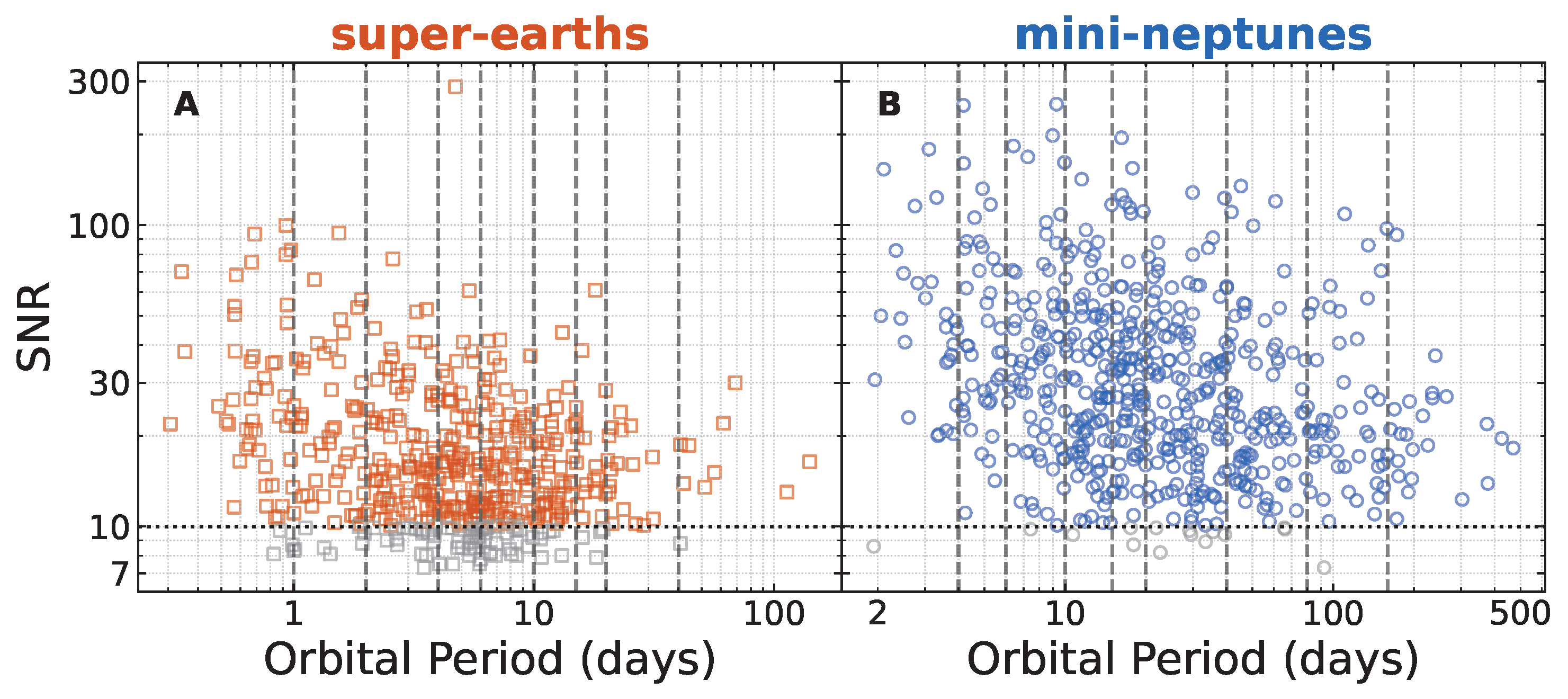}
\caption{
\textbf{The SNR as a function of orbital period.}
{\bf (A)} Orange and gray squares show the period--SNR distributions of SEs in the Kepler fiducial sample with SNR $>10$ and SNR $<10$, respectively.
The dashed vertical lines show the bin edges, which are the same as those applied in our fiducial Kepler sample.
The dotted horizontal lines indicate the SNR thresholds.
{\bf (B)} Same as panel (A), bur for MNs.
}
\label{fs16}
\end{figure}

\begin{figure}[!ht]
\centering
\includegraphics[width=1\textwidth]{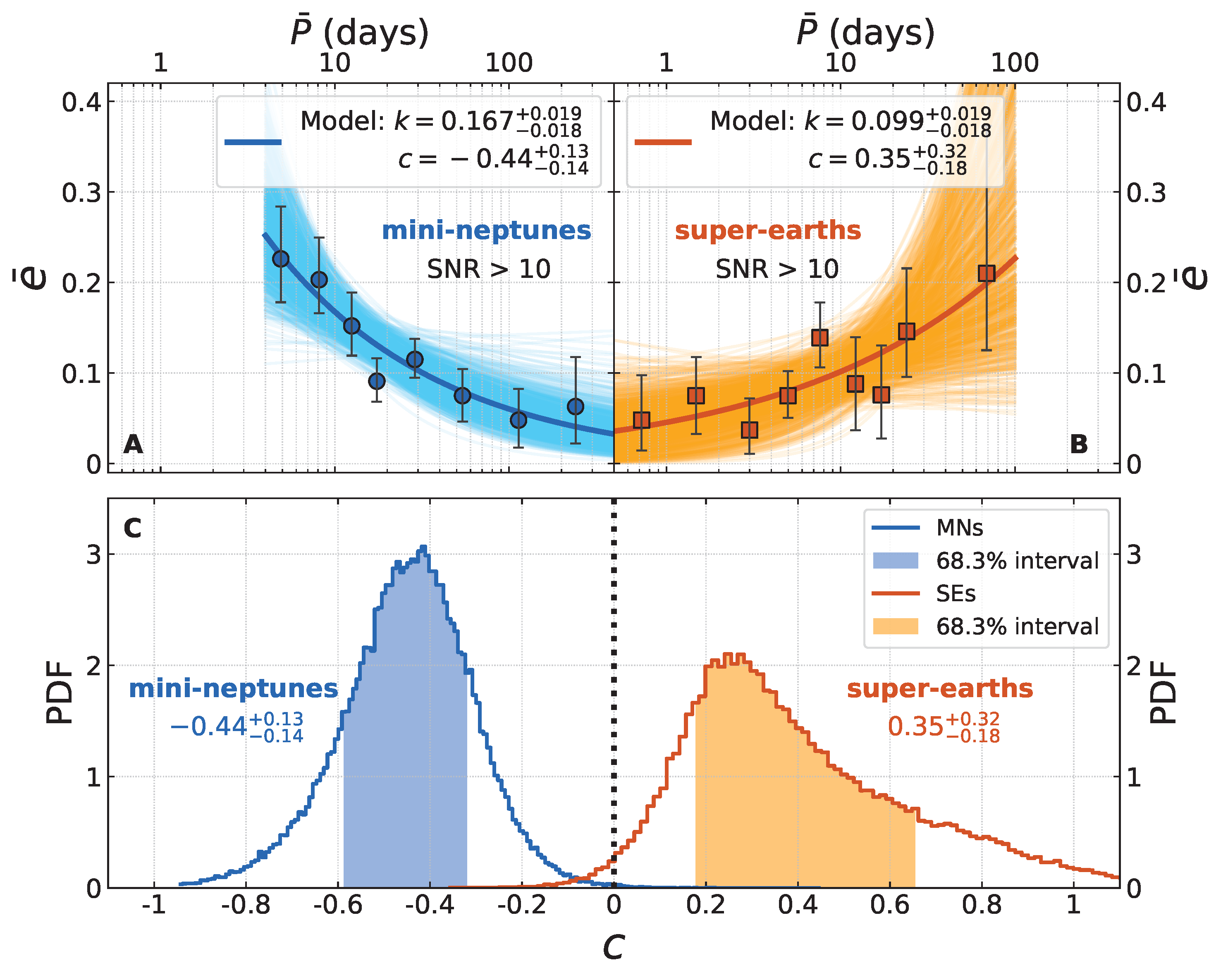}
\caption{
\textbf{Same as Fig.~\ref{f3}A, C\&D, but for the Kepler sample with SNR > 10.}
}
\label{fs17}
\end{figure}

\noindent We apply an SNR cut of SNR>7.1\cite{2011ApJ...736...19B} to exclude false positives and ensure the reliability of the measured planetary parameters (e.g. transit duration) in our fiducial Kepler sample.
Applying such a cut might exclude grazing transits or have varying effects on exoplanets with different impact parameters.
To test the effect of the SNR cut, we construct another sample with SNR>10 (Fig.~\ref{fs16}) and re-perform the fitting procedures to this sample.
Fig.~\ref{fs17} shows the resulting $P$--$e$ relations of SEs and MNs with SNR>10 are consistent with Fig.~\ref{f3}.
\end{multicols}

\clearpage

\begin{multicols}{2}
\subsection*{\hypertarget{s8}{Effect of observational bias}}
\subsubsection*{\hypertarget{s8.1}{TTV systems and MMR systems}}
\begin{figure}[!ht]
\centering
\includegraphics[width=0.95\textwidth]{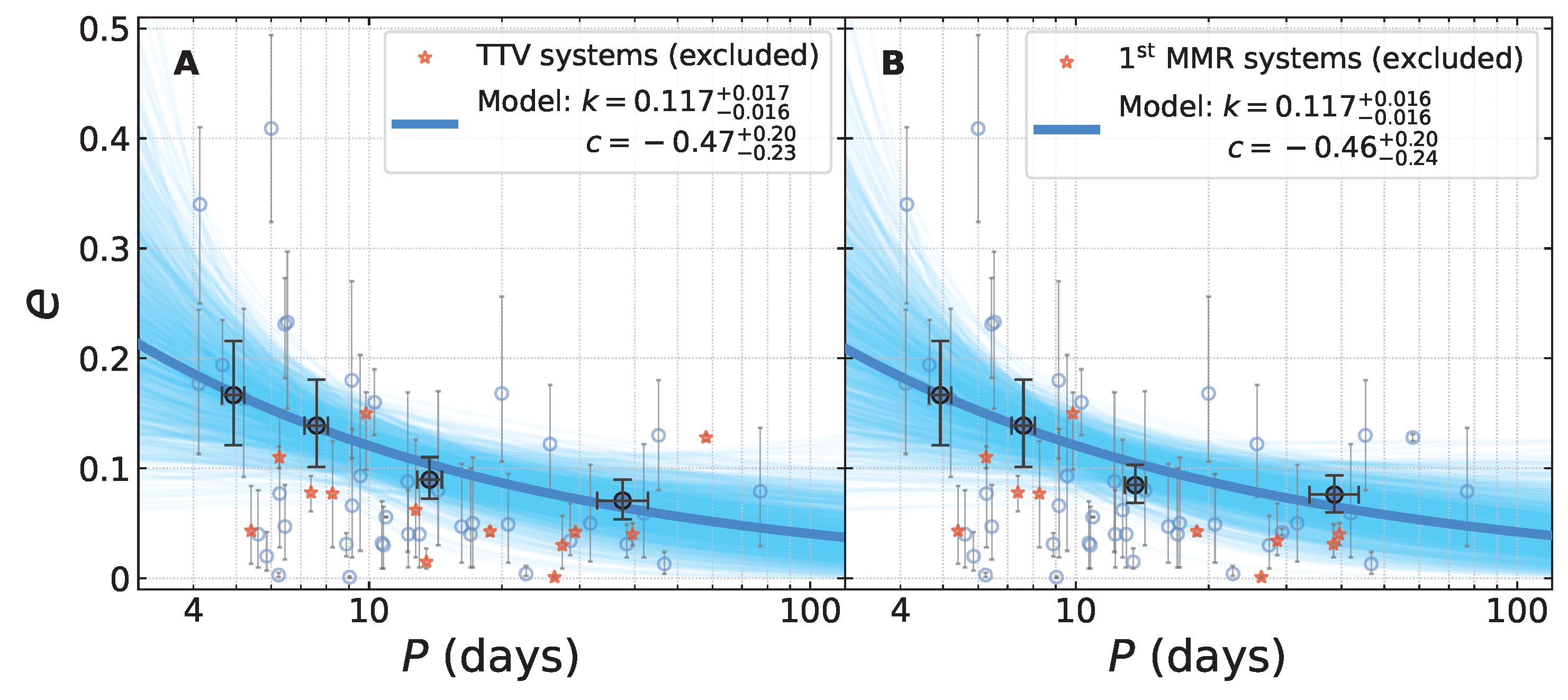}
\caption{
\textbf{Same as Fig.~\ref{f3}B, but for NEA samples with TTV or MMR systems excluded.}
{\bf (A)} Red stars are TTV systems, which are excluded during the fitting procedures.
{\bf (B)} Red stars are first order MMR systems, which are excluded during the fitting procedures.
}
\label{fs18}
\end{figure}

\noindent The exoplanets in our comparison NEA samples were detected and characterized using transit, TTV, and RV methods.
None of these three methods are strongly biased towards or away from moderate eccentric exoplanets.
However, exoplanet systems with orbital eccentricities derived solely from TTV methods are biased towards resonant (or near-resonant) systems.
These TTV systems might have a different eccentricity distribution compared to other exoplanets in the comparison NEA sample.
To test for this effect, we remove all TTV systems (13 exoplanets in total, Fig.~\ref{fs18}A) in our comparison NEA sample and re-perform the fitting procedures.
Fig.~\ref{fs18}A shows the resulting $P$--$e$ relation of MNs is consistent with Fig.~\ref{f3}.

Furthermore, we identify 10 exoplanets in first order mean motion resonance with $-0.015\leq \delta \leq +0.03$ (Fig.~\ref{fs18}B), where $\delta$ is the distant to resonance\cite{2024AJ....168..239D}:
\begin{equation}
\label{es27}
{\delta = \frac{P_{\rm out}}{P_{\rm in}} \frac{q-1}{q}-1},
\end{equation}
where $q$ is an integer ($1<q<7$), $P_{\rm out}$ is the observed orbital periods of the inner exoplanet in a neighboring pair, and $P_{\rm in}$ is the observed orbital periods of the outer exoplanet in the same pair.
We also re-performed the fitting procedures after removing MMR systems.
Fig.~\ref{fs18}B shows the resulting $P$--$e$ relation of MNs remains consistent with Fig.~\ref{f3}.

Therefore, we conclude that TTV and MMR systems do not substantially influence the measured $P$--$e$ anti-correlation of MNs in the NEA sample.
\end{multicols}

\begin{multicols}{2}
\subsubsection*{\hypertarget{s8.2}{Effect of using single-transit systems}}
\noindent The fiducial Kepler single-transit exoplanet sample are more homogeneous than the comparison NEA samples.
Compared to the NEA samples, which consist of exoplanets with orbital eccentricities measured by RV and TTV methods, our Kepler samples rely on the TDR method to uniformly derive the mean eccentricities of exoplanets in each bin.
However, using the $P$--$e$ relation of single-transit exoplanet systems to represent the entire population might introduce biases.
To investigate this, we perform a forward modeling test. 

We generate one million simulated exoplanet systems assuming eccentricities are independent of orbital periods.
Specifically, we assign $R_\star$, $M_\star$, $N_{\rm p}$, $\omega$, $e$, $i$, $a$ and $P$ to each generated systems, where $N_{\rm p}$ is the number of exoplanets.
The radii and masses of the host stars are fixed to the solar values ($R_\star=1~R_\odot$ and $M_\star=1~M_\odot$).
$N_{\rm p}$ of each system is drawn from a Poisson distribution with a mean value of $\bar{n} = 3$\cite{2018ApJ...860..101Z,2020AJ....159..164Y}. 
$\omega$ is drawn from a uniform distribution between 0 and $2\pi$.
$e$ is drawn from a uniform distribution  between 0 and 0.4.
Following previous work \cite{2016PNAS..11311431X}, we set the orbital inclinations of exoplanets (relative to the system invariant plane) to be equal to their eccentricities ($i=e$).
For $a$ and $P$ of exoplanets, we apply previous method\cite{2012AJ....143...94T,2018ApJ...860..101Z} to randomly draw $\xi\equiv R_\star/a$ according to the PDF:
\begin{equation}
\label{es28}
{\rm PDF(\xi)} = \frac{0.36}{\xi}\frac{\left(\frac{\xi}{0.074}\right)^{0.04}}{1+\left(\frac{\xi}{0.074}\right)^{3.18}},
\end{equation}
for $0.004<\xi<0.6$.
This PDF accounts for the correction of geometric transit probability\cite{2012AJ....143...94T,2018ApJ...860..101Z}.
We then obtain $a$ and $P$ from $\xi$.
\end{multicols}

\begin{figure}[!ht]
\centering
\includegraphics[width=1\textwidth]{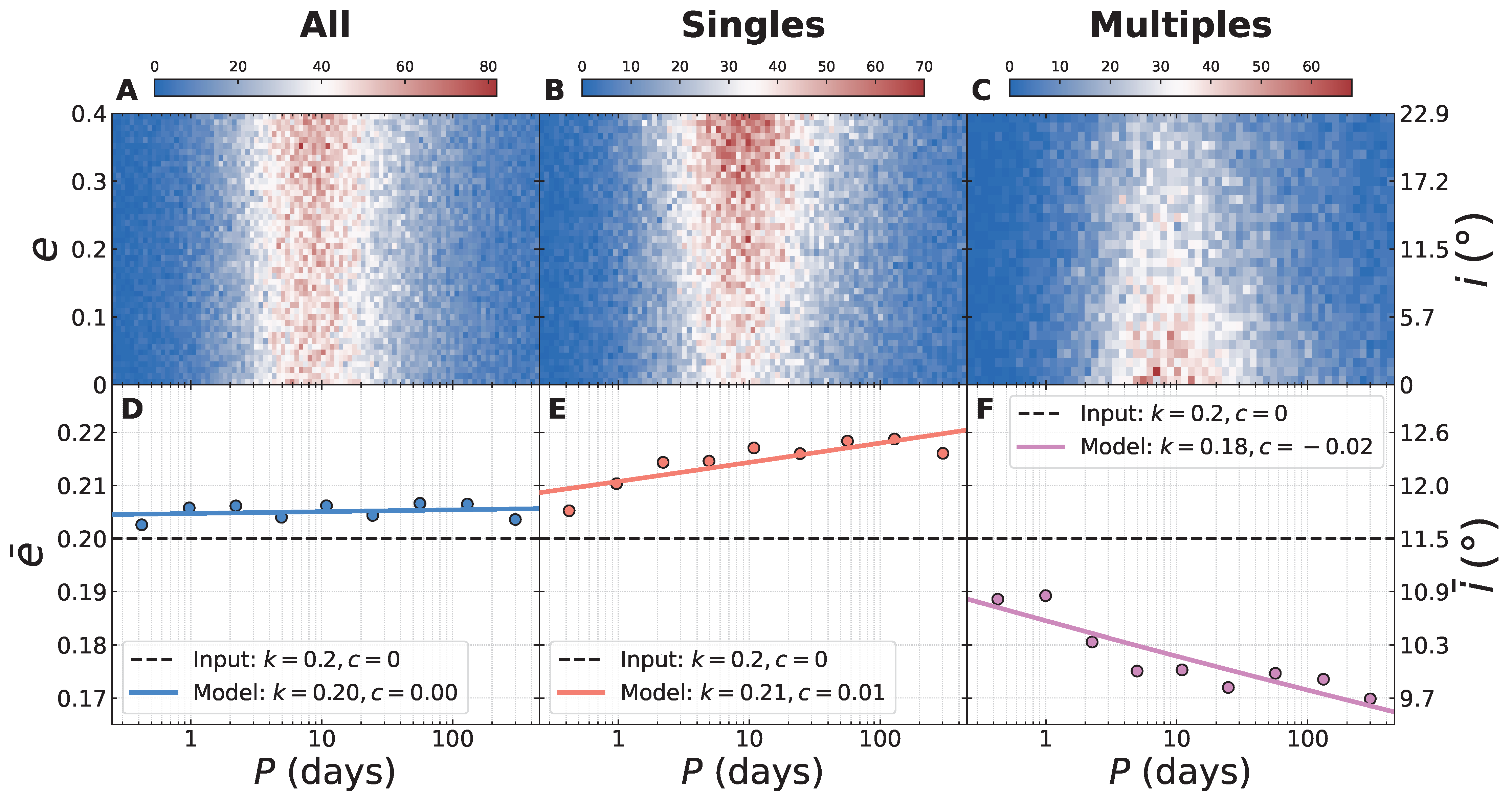}
\caption{
\textbf{The period--eccentricity distributions of simulated exoplanet systems.}
{\bf (A)} 2-D histograms of $P$--$e$ ($i$) distribution of all simulated systems, the number of exoplanets in each bin is indicated by the color bar.
{\bf (B)} Same as panel (A), but for simulated single-transit systems.
{\bf (C)} Same as panel (A), but for simulated multiple-transit systems.
{\bf (D)} The circles show $\bar{P}$-$\bar{e}$ ($\bar{i}$) distributions of all observable simulated systems in nine orbital period bins.
Colored line is a power-law model: $\bar{e} = k\cdot(\bar{P}/(10~{\rm days)})^c$ fitted to the data, which has parameters labeled in the legend.
The black dashed lines indicate the input $P$--$e$ relations ($e=0.2$ and with no correlation with orbital period) of these simulated systems.
{\bf (E)} Same as panel (D), but for simulated single-transit systems.
{\bf (F)} Same as panel (D), but for simulated multiple-transit systems.
}
\label{fs19}
\end{figure}

\begin{multicols}{2}
After the generation, we conduct a simple simulation of transit observation to the generated exoplanet systems to acquire single-transit and multi-transit exoplanet samples.
Specifically, we calculate $b$ of each generated exoplanet to determine whether it would be observed.
Assuming the distribution of $I$ (inclination of the system invariable plane relative to the observer) is isotropic (${\rm cos}I$ follows a uniform distribution between -1 and 1), the exoplanet inclination relative to the observer $I_{\rm p}$ is given by\cite{2018ApJ...860..101Z}:
\begin{equation}
\label{es29}
I_{\rm p} = {\rm arccos}({\rm cos}I\,{\rm cos}i-{\rm sin}I\,{\rm sin}i\,{\rm cos}\phi),
\end{equation}
where $\phi$ is the longitude of ascending node of the exoplanet (also termed the phase angle, which represents the orientation of the orbital plane of the exoplanet), which we randomly drew from a uniform distribution between 0 and $2\pi$.
Combining Eq.~\ref{es29} with Eq.~\ref{es3}, we obtain $b$ for each generated system, and the exoplanets with $|b|<=1$ are considered potentially observable.
We then record the orbital periods and eccentricities of these exoplanets, and categorize their systems as single-transit or multi-transit.
The $P$--$e$ distribution of all the observable simulated systems is shown in Fig.~\ref{fs19}A, single-transit exoplanets (singles) in Fig.~\ref{fs19}B and multiple-transit exoplanets (multiples) in Fig.~\ref{fs19}C.
We find singles tend to have higher eccentricities, while multiples tend to have lower eccentricities, consistent with previous work\cite{2015ApJ...808..126V,2016PNAS..11311431X,2019AJ....157...61V}.
To investigate the $P$--$e$ relations of singles and multiples, we divide these systems into nine bins according to their orbital periods (similar to our fiducial Kepler sample).
We calculate the mean orbital periods and mean eccentricities in each bin, and fitted a power-law model same as our fiducial Kepler sample to their $\bar{P}$--$\bar{e}$ distributions.
Fig.~\ref{fs19}D-F show that the $\bar{P}$--$\bar{e}$ relation of our simulated transits resembles the input intrinsic relation (with $k=0.2$ and $c=0$).
However, the singles tend to show a positive $\bar{P}$--$\bar{e}$ correlation and the multiples tend to show an anti-correlation between $\bar{P}$ and $\bar{e}$.
This is because lower orbital inclinations (eccentricities) are required for exoplanets with longer periods in multiple-transit systems to be observable.

The $\bar{P}$--$\bar{e}$ relations induced by this observational bias are at least an order of magnitude smaller than what our observed ones.
The increase/decrease of orbital eccentricities from 1 to 100 days is $\sim 0.02$, which is ten times smaller than we observed.
Therefore, the variation in the power-law index $c$ induced by the observational bias is negligible compared to the $1~\sigma$ uncertainties in our results for both SEs and MNs.

\subsection*{\hypertarget{s9}{Effect of stellar metallicity}}
\noindent For small exoplanets, there is an anti-correlation between orbital period and stellar metallicity\cite{2013A&A...560A..51A,2016AJ....152..187M}, and a positive correlation between stellar metallicity and orbital eccentricity\cite{2019AJ....157..198M,2023AJ....165..125A}.
The combination of these two relations could induce an anti-correlation between orbital period and eccentricity.
To investigate this effect, we utilize the Kepler-LAMOST small single-transit planet sample from previous work\cite{2023AJ....165..125A}.
This sample contain precise and homogeneous stellar metallicity measurements from LAMOST data release 8\cite{2012RAA....12.1197C} for small single-transit Kepler planets with $R<4R_\oplus$.
We obtain a period--metallicity relation from this sample:
\begin{equation}
\label{es30}
{\rm [Fe/H]} = -0.028~{\rm log}_{10}\left(\frac{P}{10\;{\rm days}}\right)-0.020, \\
\end{equation}
where [Fe/H] is stellar metallicity.
Eq.~\ref{es30} shows an anti-correlations between stellar metallicities and orbital periods, consistent with previous works\cite{2013A&A...560A..51A,2016AJ....152..187M}.
Because the orbital eccentricities of small exoplanets are positively correlated with the metallicities of their host star\cite{2019AJ....157..198M,2023AJ....165..125A}, this anti-correlation between planetary orbital period and stellar metallicity will induce a $P$--$e$ anti-correlations.
This projected $P$--$e$ anti-correlation could potentially weaken the observed positive $P$--$e$ correlation of SEs and strengthen the observed $P$--$e$ anti-correlation of MNs.

To remove the influence of the stellar metallicities, we utilize the metallicity--eccentricity relation for single-transit small exoplanets found in previous work\cite{2023AJ....165..125A}: $\bar{e}\propto10^{1.78\cdot{\rm \overline{[Fe/H]}}}$ (which is derived using the same Kepler-LAMOST sample mentioned above).
Based on this metallicity--eccentricity relation, we calculate corrected mean eccentricities:
\begin{equation}
\label{es31}
\bar{e}' = \bar{e} \cdot 10^{-1.78\cdot{\rm \overline{[Fe/H]}}},
\end{equation}
where $\bar{e}$ and $\overline{\rm [Fe/H]}$ are the mean eccentricities (before correction) and mean metallicities (from Eq.~\ref{es30}) in each bin, respectively.
$\overline{\rm [Fe/H]}>0$ corresponds to $\bar{e}'<\bar{e}$, and vice versa.

\begin{figure}[!ht]
\centering
\includegraphics[width=1\textwidth]{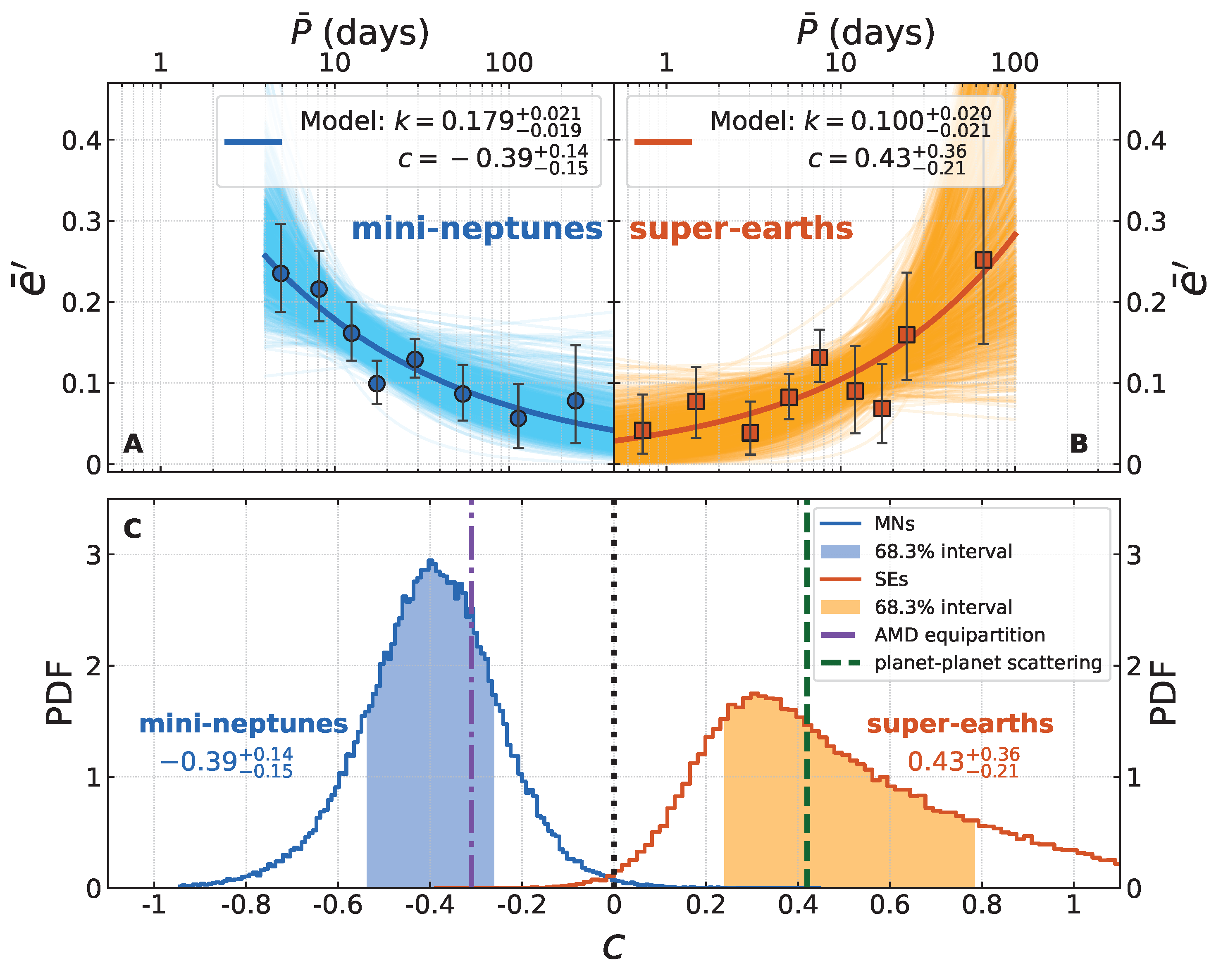}
\caption{
\textbf{Same as Fig.~\ref{f3}A, C\&D, but for the Kepler sample with metallicity correction.}
The purple dot-dashed line indicates the power-law index predicted by AMD equipartition (Eq.~\ref{es22}).
The green dashed line indicates the power-law index predicted by PPS (Eq.~\ref{es17}).
}
\label{fs20}
\end{figure}

After correcting the effect of stellar metallicity, we perform the same fitting procedures as our fiducial Kepler sample to the $\bar{P}-\bar{e}'$ distributions of SEs and MNs (Fig.~\ref{fs20}.)
For SEs, the positive $P$--$e$ correlation after metallicity correction is stronger (a larger power-law index of $c=0.43$ and a lower p-value of $7.1\times13^{-3}$) than was measured for our fiducial Kepler sample.
For MNs, the $P$--$e$ anti-correlation after metallicity correction becomes weaker (a larger index of $c=-0.39$) than was measured for our fiducial Kepler sample, but the p-value $4.6\times13^{-3}$ remains small (corresponding to a significance level of $2.8~\sigma$).
For both SEs and MNs, the power-law indices $c$ after the metallicity correction are closer to the values of theoretical predictions (Fig.~\ref{fs20}C).

We conclude that the combination of period--metallicity and metallicity--eccentricity relations can partially explain the observed $P$--$e$ anti-correlation for MNs, and potentially weakens the observed positive $P$--$e$ correlation of SEs.
After correcting for this effect, the positive $P$--$e$ relation of SEs strengthens, and the $P$--$e$ anti-correlation of MNs weakens.
Nevertheless, the $P$--$e$ relations remain different between SEs and MNs after the correction: the posterior probability distributions of $c$ for SEs and MNs (Fig.~\ref{fs20}C) overlap by a fraction of $2.9\times13^{-3}$, corresponding to a significance level of $\sim 3.6~\sigma$.
\end{multicols}

\begin{multicols}{2}
\subsection*{\hypertarget{s10}{Previous studies}}
\noindent Previous works have studied the eccentricity distribution of small exoplanets\cite{2015ApJ...808..126V,2019AJ....157...61V,2020A&A...635A..37C,2024AJ....168..115B}, however, none of them reported the relations between orbital periods and eccentricities that we found in this work.
Here we consider why our results differ from theirs

\begin{figure}[!ht]
\centering
\includegraphics[width=0.8\textwidth]{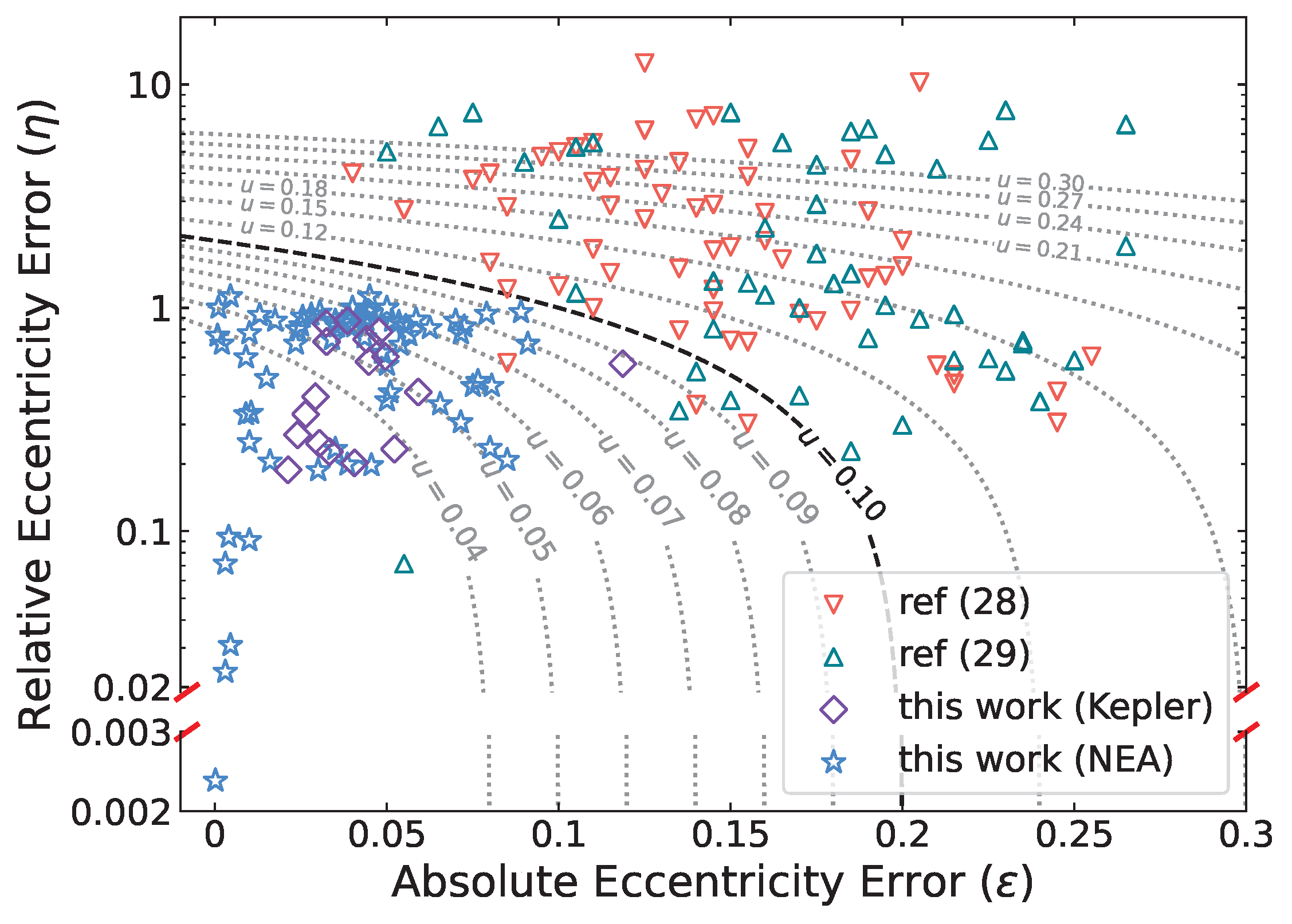}
\caption{
\textbf{The eccentricity uncertainty from different previous works.}
Eccentricity uncertainties in this work (purple diamonds show fiducial Kepler and blue stars show comparison NEA sample) compared to those in previous work as orange downward triangles\cite{2015ApJ...808..126V} and green upwards triangles\cite{2019AJ....157...61V}.
The gray dotted contours show combined uncertainties $u$ (Eq.~\ref{es9}), and the black dash contour indicates the maximum combined uncertainties ($u=0.10$) of this work.
}
\label{fs21}
\end{figure}

Our samples have smaller eccentricity uncertainties compared to previous works.
Previous studies\cite{2015ApJ...808..126V,2019AJ....157...61V} investigated the orbital eccentricities of small Kepler exoplanets using the eccentricities of individual exoplanets, which have large uncertainties.
Our TDR method instead investigates the mean eccentricities of a group of exoplanets (in each bin), which reduces the uncertainties. 
Fig.~\ref{fs21} shows our fiducial Kepler and comparison NEA samples have smaller uncertainties ($u \equiv 0.5\varepsilon + 0.05<0.1$).
Most of the exoplanets in previous studies have larger eccentricity uncertainties.
As discussed above, the $P$--$e$ relations we found would be obscured by the those higher uncertainties.

Previous studies did not separate their samples into SEs and MNs: they either investigated the small exoplanets with $R<6~R_\oplus$\cite{2015ApJ...808..126V,2019AJ....157...61V}, small exoplanets with $R<3~R_\oplus$\cite{2020A&A...635A..37C}, or low-mass exoplanets with $m<0.165~M_{\rm J}$\cite{2024AJ....168..115B}.
Because we find that SEs and MNs follow opposite $P$--$e$ relations, mixing these two populations could produce a flat eccentricity-period distribution as they found.
\end{multicols}

\end{document}